\documentclass{aa}  

\usepackage{graphicx}
\usepackage{txfonts}
\usepackage{amsmath}
\usepackage{natbib}

\begin{document}

   \title{SWEET-Cat updated \thanks{Based on observations collected at the European Organisation for Astronomical Research in the Southern Hemisphere under ESO programs 096.C-0092, 097.C-0280, 098.C-0151, and data obtained from the ESO Science Archive Facility under several requests (request numbers 273579-274977).} \thanks{Tables A.1-A.3 are also available in electronic form at the CDS via anonymous ftp to cdsarc.u-strasbg.fr (130.79.128.5) or via http://cdsweb.u-strasbg.fr/cgi-bin/qcat?J/A+A/} }

   \subtitle{New homogenous spectroscopic parameters}

   \author{S. G. Sousa\inst{1}                  
          \and V. Adibekyan\inst{1}             
          \and E. Delgado-Mena\inst{1}
          \and N. C. Santos\inst{1,}\inst{2}
          \and D. T. Andreasen\inst{1}             
          \and A. C. S. Ferreira\inst{1}
          \and M. Tsantaki\inst{1}              
          \and S. C. C. Barros\inst{1}          
          \and O. Demangeon\inst{1}
          \and G. Israelian\inst{3}
          \and J. P. Faria\inst{1}
          \and P. Figueira\inst{4,}\inst{1}
          \and A. Mortier\inst{5}               
          \and I. Brand\~ao\inst{1}
          \and M. Montalto\inst{6}
          \and B. Rojas-Ayala\inst{7}
          \and A. Santerne\inst{8,}\inst{1}
          }

          \institute{Instituto de Astrof\'isica e Ci\^encias do Espa\c{c}o, Universidade do Porto, CAUP, Rua das Estrelas, 4150-762 Porto, Portugal
          \and Departamento de F\'isica e Astronomia, Faculdade de Ci\^encias, Universidade do Porto, Rua do Campo Alegre, 4169-007 Porto, Portugal
          \and Instituto de Astrof\'isica de Canarias, 38200 La Laguna, Tenerife, Spain
          \and European Southern Observatory, Alonso de Cordova 3107, Vitacura, Santiago, Chile
          \and Centre for Exoplanet Science, SUPA, School of Physics and Astronomy, University of St Andrews, St Andrews KY16 9SS, UK
          \and Dipartimento di Fisica e Astronomia "Galileo Galilei", Universit\'a di Padova, Vicolo dell'Osservatorio 3, Padova IT-35122, Italy
          \and Departamento de Ciencias Fisicas, Universidad Andres Bello, Fernandez Concha 700, Las Condes, Santiago, Chile
          \and Aix Marseille Univ, CNRS, CNES, LAM, Marseille, France
}

   \date{Received September 15, 1996; accepted March 16, 1997}

 
  \abstract
   {Exoplanets have now been proven to be very common. The number of its 
   detections continues to grow following the development 
   of better instruments and missions. One key step for the understanding of 
   these worlds is their characterization, which mostly depend on their host stars.}
   {We perform a significant update of 
   the Stars With ExoplanETs CATalog (SWEET-Cat), a unique compilation of 
   precise stellar parameters for planet-host stars provided for the exoplanet
   community.
   }
   {We made use of high-resolution spectra for planet-host stars, either observed
   by our team or found in several public archives. The new spectroscopic parameters 
   were derived for the spectra following the same homogeneous process (ARES+MOOG). 
   The host star parameters were then merged together with the planet properties 
   listed in exoplanet.eu to perform simple data analysis.
   }
   {We present new spectroscopic homogeneous parameters for 106 planet-host stars. Sixty-three 
   planet hosts are also reviewed with new parameters. We also show that there is a 
   good agreement between stellar parameters derived for the same star but using 
   spectra obtained from different spectrographs. The planet-metallicity correlation 
   is reviewed showing that the metallicity distribution of stars hosting low-mass 
   planets (below 30 M$_{\oplus}$) is indistinguishable from that from the solar 
   neighborhood sample in terms of metallicity distribution.}
  {}


   \maketitle
%

\section{Introduction}

The first extrasolar planets around solar-type stars to be discovered, hot jupiters,
have completely changed our understanding of planet formation and evolution based on 
the solar system alone \citep[e.g.,][]{Mayor-1995}. The surprises were far from over 
and in recent years, with the detection of smaller and lower mass planets, and our 
theoretical view continues to be shaped by the observational evidence. Today $\sim$ 
3800 extrasolar planets have been discovered orbiting $\sim
$ 2800 solar-type stars \citep[exoplanet.eu][]{Schneider-2011}. The higher 
numbers of discoveries allows us to perform better statistical studies on the 
different types of planetary systems. 

Although today we are convinced that extrasolar planets are present around almost 
every (dwarf or main sequence) star \citep[e.g.,][]{Mayor-2011}, the first 
observational evidence, based on the first massive planets detections, showed that 
planets were more frequently orbiting metal-rich stars \cite[e.g.,][]{Santos-2004b, 
Valenti-2005}. Soon after, it was reported that the metallicity correlation was not 
clear for the first detections of lower mass and small planets \citep[e.g.,][]{Sousa-2008, 
Ghezzi-2010, Schlaufman-2011, Buchhave-2012, Wang-2015, Buchhave-2015}. Other 
observational evidence has been reported involving the host stellar properties and 
the properties of planets. Correlations such as the planetary radius versus metallicity 
\citep[e.g.,][]{Buchhave-2014, Schlaufman-2015}, planet-period versus metallicity, or 
eccentricity versus metallicity \citep[e.g.,][]{Beauge-2013, Adibekyan-2013, Dawson-2015, 
Mulders-2016, Wilson-2018}, all show interesting evidence that is important for the 
understanding of planet formation. It is therefore crucial not only to increase the 
number of discoveries but also to characterize planetary systems precisely since 
they play a fundamental role in the interpretation of these results.

Precise and accurate fundamental planetary parameters (mass, radius, and mean 
density) are needed to distinguish between solid rocky, water rich, or otherwise gas 
dominated planets. To achieve this we need a precision of planetary mass up to 10\% 
and a radius up to 5\% \citep[or even 2\% for further bulk characterization; e.g.,][]
{Wagner-2011, Bean-2011}. The derivation and precision of the planetary properties 
depends considerably on the deduced parameters of the stellar 
host \citep[e.g.,][]{Bouchy-2004, Torres-2012, Mortier-2013b, Sousa-2015a}. 
Interestingly, it has been shown that the quality of the spectroscopic and 
photometric observations affects the properties of the planets more significantly 
than the model-dependent systematics, such as using different stellar evolution models 
for the determination of stellar mass and radius \citep[e.g.,][]{Southworth-2009}. It 
is therefore extremely important to use high-quality data to refine the values for 
these stellar properties in order to obtain more precise and accurate stellar masses 
and radii and thus more precise and accurate planetary masses and radii. Furthermore, 
to minimize global statistical errors, a uniform analysis is 
required \citep[see, e.g.,][]{Torres-2012} to guarantee homogeneity of the results. 
Using different methods to derive stellar properties leads to discrepancies in the results which, in turn, deteriorates the significance of the statistical analyses.

For this reason a great effort has been made in the creation and 
continuous update of the currently largest homogeneous catalog of stellar parameters 
for planet hosts (SWEET-Cat\footnote{www.astro.up.pt/resources/sweet-cat}) first 
presented in \citet[][]{Santos-2013}, and then updated with new parameters for 
a significant number of planet hosts in \citet[][]{Sousa-2015a} and in 
\citet[][]{Andreasen-2017}. In this work we continue the series 
of SWEET-Cat papers using our uniform method, which has been tested and improved upon 
extensively. We collected spectra from different spectrographs to derive new 
homogeneous stellar parameters for 106 planet hosts. We then use the updated 
catalog to review the metallicity correlations and make these parameters available for 
the community; these new values can be used for further statistical studies in the quest for clues 
for the formation and evolution of the planets.

The work is divided into the following sections: Section 2 describes the spectroscopic 
data compilation for SWEET-Cat, together with some details on spectra characteristics 
that our team has compiled to date. Section 3 shows results for the new 
spectroscopic stellar parameters derived as well as a comparison with literature 
values. In addition we show that the homogenization of our method is valid with 
the use of many different high-resolution spectrographs. Section 4 presents 
some up-to-date metallicity correlations relevant for planet formation. We conclude 
in Section 5 with a summary of the work presented.

\section{Spectroscopic data}

In this section we describe the spectroscopic data used to derive stellar parameters 
in recent years. Some of this data were already relatively old (more than ten 
years), obtained from different spectrographs, and were already used in previous 
works. These were collected from observations with diverse scientific goals, mostly 
from radial-velocity (RV) planet surveys (e.g., HARPS), and other data were obtained 
specifically for planet-host spectroscopic stellar characterization (e.g., FEROS and 
UVES). To better describe all sources of the different spectroscopic data used today 
for SWEET-Cat, we divided the discussion into the following items: legacy data, recent 
observations, and public data from spectroscopic archives.

\subsection{Compiling legacy spectroscopic data}

One of the goals of this work was to compile all the spectra that our team has used 
in the past works which provides parameters for SWEET-Cat. The list of spectrographs 
used in SWEET-Cat is already relatively large and can be seen 
in Table \ref{tab_spectrographs}. We compiled all the spectra from our previous works in the literature to derive the stellar parameters. These include \citet[][]{Ammler-2009, 
Andreasen-2017, Boisse-2010, Boisse-2012, Bonomo-2012,
Damasso-2015, DelgadoMena-2014, Diaz-2012, Diaz-2016, Gillon-2007, Chew-2013, 
Hebrard-2010b, Hebrard-2016,
Lillo-Box-2016, Melo-2007, Montalto-2012, Mortier-2013a, Mortier-2013b,
Moutou-2006, Moutou-2011, Moutou-2014, Moutou-2015, Neves-2013, Neves-2014, 
Osborn-2017, Pont-2008, Rey-2017,
Sahlmann-2011, Santerne-2014, Santerne-2016, Santos-2004b, Santos-2005, Santos-2006, 
Santos-2008, Santos-2013,
Segransan-2010, Sousa-2006, Sousa-2008, Sousa-2011, Sousa-2011b, Sousa-2015a,
Tamuz-2008, Tsantaki-2013, Tsantaki-2014, Udalski-2008, Wilson-2016, daSilva-2007}.

The original data formats depend on several factors such as the instrument used or 
the pipeline used to process the data. Most of the spectra are stored in standard Flexible Image Transport System (FITS) 
files, but some were stored in text format. In order to 
homogenize all the data we made sure that all the individual spectra were stored in a 
standard 1D fit format, easy to read by standard routines and in particular by the Automatic Routine for line Equivalent widths in stellar Spectra  
\citep[ARES; see ][]{Sousa-2007, Sousa-2015b}, which requires specific FITS header keywords 
(e.g., CDELT1 and CRVAL1).

It is not our goal here to completely describe all the details of this technical 
step, but rather to provide a short description. Most of the spectra were 
already reduced and stored in this format. For the text files, the transformation is 
trivial, unless the wavelength spacing is not constant. In these cases we interpolated the spectra to derive an evenly spaced wavelength coverage, which is a 
requirement for the adopted FITS format. The 2D echelle reduced spectra 
actually needed more caution for the correct transformation into the 
standard 1D spectra. For the blaze-corrected and normalized 2D spectrum, the 
transformation is trivial and a simple merging of the orders was performed; we took special 
caution when the orders overlapped in wavelength. For these we used the 
\textit{scombine} routine in Image Reduction and Analysis Facility (IRAF) that allow us to combine different orders into a 1D 
spectrum. For rather old spectra (e.g., SARG and some old FIES observations) the 2D 
spectra are still contaminated with the blaze function. To make the situation more 
complex, in a few cases there were also some strange peak structures in the limits 
of the orders, most likely due to reduction problems. These creates difficulties for 
an automatic merging of the orders and for those cases we went 
spectra by spectra, order by order, to clean out the peak structures in the spectra 
as best as possible before the merging of the orders into a 1D spectrum.

\begin{table}[t]
\caption[]{Spectrographs used in SWEET-Cat}
\small
  \begin{center}

    \begin{tabular}{cccc}
    \hline
    \hline
    Spectrograph &   Spec range (\AA)  & R  & Observatory  \\
    \hline

    HARPS        & 3600 -  6900  & 110000 & ESO La Silla (Chile) \\
    UVES         & 4800 -  6800  & 100000 & ESO Paranal (Chile) \\
    CORALIE      & 3900 -  6800  &  50000 & ESO La Silla (Chile) \\
    ESPADONS     & 3700 - 10500  &  80000 & CHFT (USA) \\
    FEROS        & 3800 -  9200  &  48000 & ESO La Silla (Chile) \\
    SOPHIE       & 3872 -  6943  &  75000 & OHP (France) \\
    ELODIE       & 3895 -  6815  &  42000 & OHP (France) \\
    SARG         & 3700 - 10000  &  85000 & ORM (Spain)  \\
    NARVAL       & 3700 - 10500  &  80000 & TBL (France) \\
    FIES         & 3700 -  8300  &  67000 & ORM (Spain) \\
    UES          & 3600 -  7250  & 55000  & ORM (Spain) \\
    
    \hline
    \end{tabular}
  \end{center}
\label{tab_spectrographs}
\tablefoot{Many of these spectrographs have different configurations available for 
observation. We list the typical spectral range and resolution for the 
observations that we collected or found in archives.} 
\end{table}

\begin{figure}[ht]
  \centering
  \includegraphics[width=8.5cm]{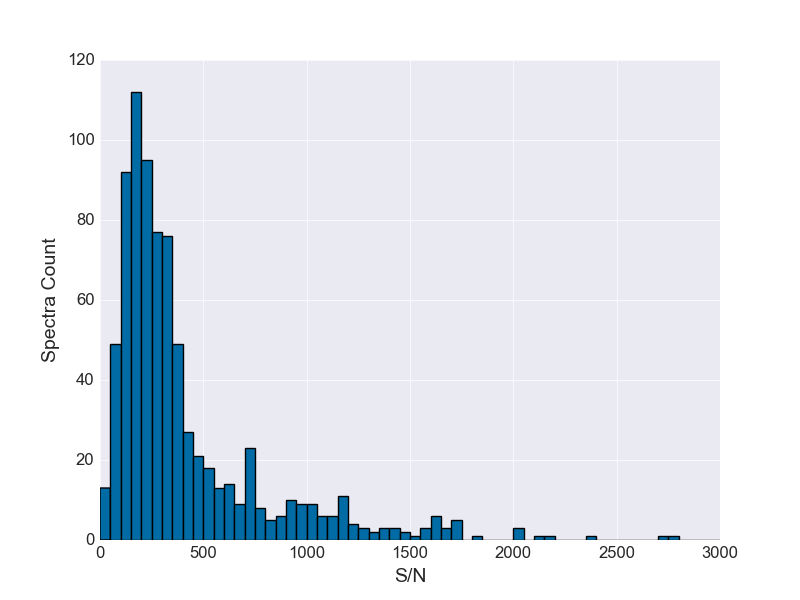} 

  \includegraphics[width=8.5cm]{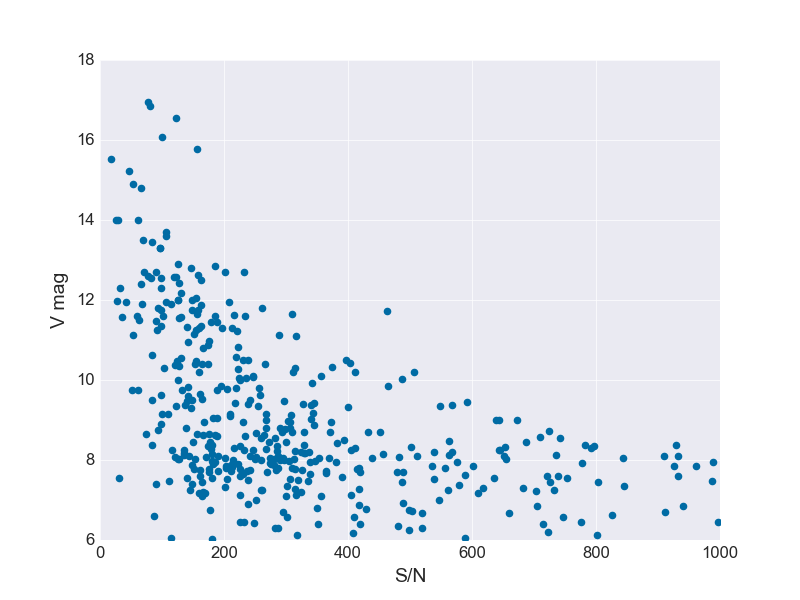}
  \caption{S/N distribution(top panel) and S/N vs. V magnitude (bottom panel) for SWEET-Cat spectral data. The bottom panel is cut at S/N of 1000 for a better visualization of the correlation.}
  \label{fig_sn}
\end{figure}

\subsection{Recent spectroscopic observations and data reduction}

During recent years several of our ESO proposals have been successfully awarded 
with which we obtained high-resolution spectra with UVES \citep[][]
{Dekker-2000} at the UT2 telescope in Paranal, Chile. The spectra that we have 
collected in recent ESO periods include programs 096.C-0092, 097.C-0280, and 
098.C-0151.

In all these proposals we took advantage of the use of Image Slicer \#3, which allows 
us to observe with poor seeing conditions and still catches most of the flux. Since we used 
a line by line analysis in our procedures, we proposed to observe with the smallest 
slit (0.3``), which gives a higher resolution for a better spectroscopic analysis. 
The typical spectra configuration used for UVES is RED580, which covers the 
optical spectrum in the interval 476-684nm; this allows us to observe many lines 
of different elements. The spectra collected in these proposals are supposed to 
achieve a S/N of 300 for the stars brighter than V=10 and S/N of 150 for the 
faintest stars. For the brightest stars single exposures are enough to reach the 
requested S/N. For the fainter stars we choose to divide the requested exposure times 
in chunks of a maximum of 1800 seconds of exposure. With this observational strategy, 
we avoid the overpopulation of cosmic rays, which becomes more difficult to correct 
for large exposure times.

The data reduction of these observations were carried out using the UVES Pipeline in 
Reflex \citep[][]{Freudling-2013} provided by ESO following the standard recipes and 
suggestions mentioned in the UVES pipeline manual to reduce UVES data observed with 
the image slicer.

The combination of the spectra was then carried out using the IRAF routine 
\textit{scombine}, which was used to combine the REDl and REDu spectra observations, 
and later this routine was used to combine these spectra into one final 1D spectrum. 

\begin{figure*}
  \centering
  \includegraphics[width=19cm]{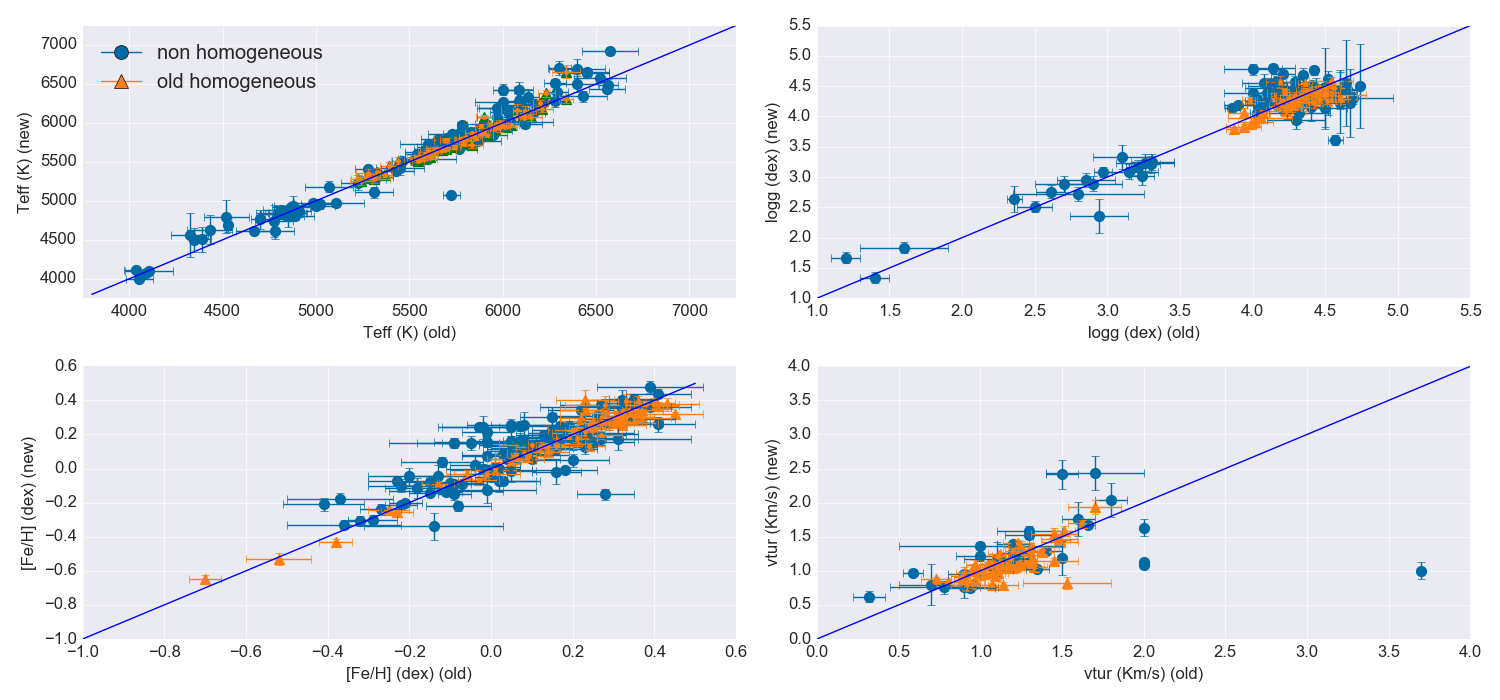}
  \caption{Comparison between old parameters listed in SWEET-Cat and the new parameters derived in this work. 
  The blue filled circles represent parameters with recent spectroscopic data. The orange triangles 
  represent new rederived parameters with legacy spectroscopic data.}
  \label{fig_param}
\end{figure*}

\begin{table}
\caption[]{Statistics of SWEET-Cat as of April 2018}
  \begin{center}

    \begin{tabular}{ll}
    \hline
    \hline
    Number     &   Description  \\
    \hline

2631 & stars in SWEET-Cat (linked from exo.eu) \\
808  & individual stellar spectra (different instruments) \\
653  & stars with spectra \\
562  & planet hosts with homogeneous parameters \\
541  & FGK stars with homogeneous parameters \\
113  & stars with spectra but no homogeneous parameters \\
106  & new parameters included in this work \\
123  & stars without spectra with V $<$ 12 \\    
    \hline
    \end{tabular}
  \end{center}
\label{tab_stats}
\end{table}

\subsection{Spectra from the public archives}

Most of the pubic data that we were able to find came from the ESO 
archive\footnote{http://archive.eso.org/eso/eso\_archive\_main.html} with the help of 
\textit{astroquery}\footnote{http://astroquery.readthedocs.io}; more specifically we 
looked for reduced data, which are also available. We looked for relatively high 
resolution and good S/N (also considering possible combinations of low S/N individual 
spectra). We selected the ESO instruments FEROS, HARPS, and UVES for 
this search in this archive. When many spectra were available for the same star, 
probably coming from RV planet search surveys, we downloaded a sufficient number of spectra to reach a 
S/N of about 2000. We note that these S/N values comes directly from photon counts and 
the real S/N should depend on other noise sources (e.g., background 
subtraction). Also, for the faint stars in the archive, when sufficient spectra were 
available, we discarded the very low S/N spectra for the combination. To give an idea 
of the amount of data collected from ESO during this process we made more than 100 
archive requests with around 34 Gb of data in total.
Other archives were also used to search for spectra for SWEET-Cat. These include 
current public archives containing spectra from SOPHIE\footnote{http://atlas.obs-
hp.fr/sophie/}, ESPADONS\footnote{http://www.cadc-ccda.hia-iha.nrc-cnrc.gc.ca/en/
cfht/}, and FIES\footnote{http://www.not.iac.es/archive/}.

\subsection{SWEET-Cat spectroscopic data statistics}

As of April 2018 we had a total of 2\,631 planet-host stars (source exoplanet.eu). We 
were able to gather spectra for 653 of these, including the new sources presented in this work. 
The reason for this relatively low number of stars with spectra is because many are relatively faint stars observed by Kepler \citep[][]
{Borucki-2010}. Given the faintness of the majority of these stars, it is clear that 
the observation and the collection of high-quality, high-resolution spectra is quite 
difficult and costly. On top of that, Kepler stars are not observable from ESO 
facilities, which are our most direct source of data. When considering bright stars, 
which is also fundamental for RV follow up and planetary mass measurements \citep[][]
{Fortier-2014, Rauer-2014}, the percentage of stars with spectra and homogeneous 
parameters becomes quite high. With the work presented in this paper, considering the addition 
of 106 new stars with homogeneous parameters, we have $\sim$90\% and 
$\sim$80\% completeness for stars with $V<9$ and $V<12,$ respectively. Completeness 
represents the number of planet hosts with homogeneous spectroscopic parameters 
relative to the number of stars with planets in SWEET-Cat.

Table \ref{tab_stats} sums up some numbers relative to the planet hosts and their 
spectra collected using different instruments as well as the number of stars with 
derived homogeneous parameters. We note that we have seven 
planet-hosts stars with low-quality spectra in the database for which we did not 
derive reliable parameters. A histogram of the S/N for all the final combined 
SWEET-Cat spectra is also presented in the top panel of Fig. \ref{fig_sn}. 
The S/N value was derived automatically using the same procedure as used in the ARES 
code and considering the recommended spectral regions \citep{Sousa-2015b}. Most of 
the very high S/N spectra come from the combination of several spectra observed 
during RV programs. The correlation between the S/N of 
our spectra with the V magnitude of the planet-host stars is also clear (see bottom panel of Fig. 
\ref{fig_sn}).


\section{Stellar parameters}

\subsection{Spectroscopic parameters and stellar masses}

The spectroscopic analysis was carried out with ''ARES+MOOG`` \citep[In this work 
we used ARES v2, and MOOG2014. For more details, see][]{Sousa-2014} both for 
the new spectroscopic data and for part of the legacy data that already had derived 
homogeneous parameters before 2008. The analysis is based on the excitation and 
ionization balance of iron abundance, where the absorption lines are consistently 
measured with the ARES code \citep[][]{Sousa-2007, Sousa-2015b} 
and the abundances are derived in local thermodynamic equilibrium (LTE) with the MOOG code \citep[][]{Sneden-1973} and 
uses a grid of Kurucz ATLAS9 plane-parallel model atmospheres \citep[][]
{Kurucz-1993}. The same methodology can also be applied with FASMA\footnote{http://
www.iastro.pt/fasma/}\citep{Andreasen-2017}. The method has been applied in our 
previous spectroscopic studies of planet hosts \citep[e.g.,][]{Sousa-2008, Sousa-2011, 
Mortier-2013b, Sousa-2015a}. 

The line list used here was that from \citet[][]{Sousa-2008} except 
for the stars whose effective temperature is below 5200 K ($T_{eff} < 5200 K$). For 
these stars we rederived parameters using a more adequate line list for cooler stars, 
which was compiled by \citet[][]{Tsantaki-2013}. The atomic data, log gfs, were recalibrated in the same way as in previous works, but in this work we used MOOG2014. There are small changes in log gf values, which most likely come from numerical differences from a different 
compilation of the code.

\begin{figure*}
  \centering
  \includegraphics[width=19cm]{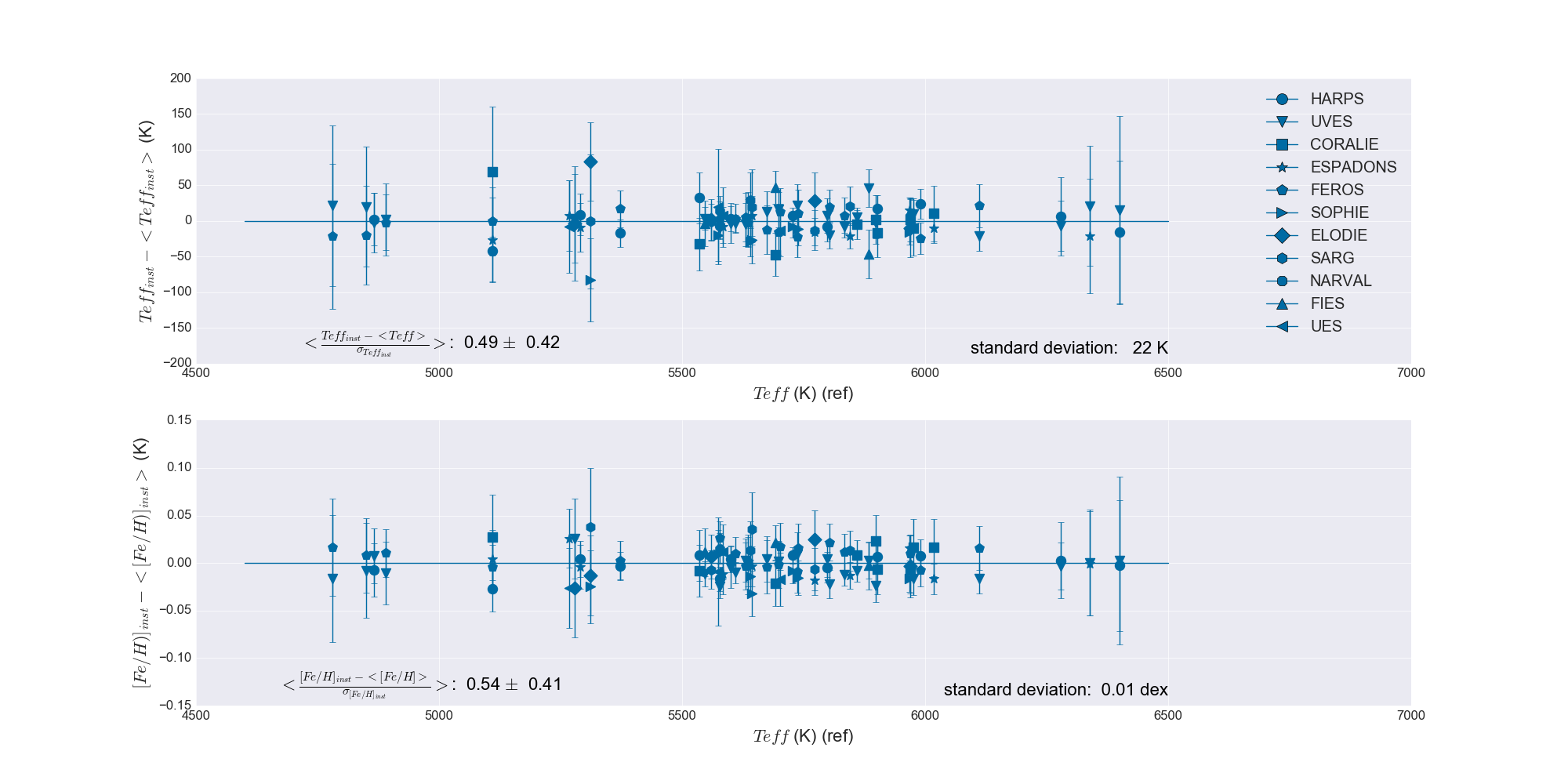}
  \caption{Comparison between parameters derived from different spectrographs with ARES+MOOG.}
  \label{fig_inst}
\end{figure*}

The stellar masses presented in this work were derived with the calibration presented in 
\citet[][]{Torres-2010} and a correction presented in \citet[][]
{Santos-2013} was applied for the cases in which the calibration gives values between 0.7 and 
1.3 $M_\odot$. The errors for these mass values are computed as in \citet[][]
{Santos-2013}, where for each case 10\,000 random values of effective temperature, 
surface gravity, and stellar metallicity were drawn from a Gaussian distribution. We used the 
uncertainties of the spectroscopic parameters this way to estimate the 
distribution of the derived mass. The error for the mass is extracted from the 
standard deviation of this mass distribution for each star.

Table \ref{tab_parameters} presents the new parameters derived with ARES+MOOG in this 
work for the new spectra collected. In this table we report the spectroscopic 
parameters (effective temperature (T$_{\mathrm{eff}}$), surface gravity ($\log{g}
_{spec}$), microturbulance ($\xi_{\mathrm{t}}$), and iron abundance ([Fe/H] - used as 
a proxy for metallicity), the number of iron lines used in the spectroscopic 
analysis, and the stellar mass. This table also presents the instrument and a column 
identifying which parameters are adopted for the planet hosts in the SWEET-Cat online table. 

We note that in \citet[][]{Sousa-2008} we also used HIgh-Precision PARallax COllecting Satellite (HIPPARCOS) parallaxes to 
derive surface gravities, and in that work we observed some differences between the  
spectroscopic and trigonometric $\log{g}$. However, although the 
spectroscopic surface gravity 
is the less constrained parameter in our analysis, the positive side is that the 
other parameters (T$_{\mathrm{eff}}$ and [Fe/H]) are almost independent 
of the $\log{g}$. This means that when fixing the $\log{g}$ to a different 
value derived with the use of parallaxes, 
interferometry, and asteroseismology we do not see any significant changes in the other 
parameters. In the online table the $\log{g}$ values listed are those directly 
derived from our spectroscopic analysis and are not corrected for known systematics, 
but we suggest the user employ the corrections discussed in \citet[][]{Mortier-2014}. 

There are also a few stars for which the application of an equivalent width, such as 
the ARES+MOOG method, is not appropriate owing to strong blends in the spectra. Some of 
these stars reveal a relatively high stellar rotation ($\gtrsim$ 10-15 Km/s) and 
others are quite cool stars (T$_{\mathrm{eff}} \lesssim $ 4200 K). For these stars we 
applied a method based on synthetic spectra, which preserves the 
homogeneity; this method is described in \citet[][]{Tsantaki-2014}, in which we have shown that the parameters derived in this way are in the same scale as the parameters derived by our Equivalent Width (EW) method. The parameters for these 11 stars are presented in Table 
\ref{tab_parameters_syn}. For these stars we were also able to derive the rotational 
velocity ($v \sin(i)$).

In SWEET-Cat there are a significant amount of planet hosts, for which the 
spectroscopic parameters were derived more than ten years ago. These represent some 
of the first planets discovered, many of which are hot jupiters for which the 
spectroscopic analysis was based on a smaller line list of iron lines \citep[e.g.,][]
{Santos-2004b}. We have shown that these parameters are 
consistent and in the same scale as those derived with the current largest line 
lists \citep[][]{Sousa-2008, Tsantaki-2013}. Consequently, we took the opportunity to run our codes 
with these line lists for the whole sample to reanalyze these stars with the current 
method, which gives more statistical strength to the results. Therefore, all the 
spectroscopic parameters that were derived with this small line list were rederived 
in this work. Table \ref{tab_parameters_old} presents the results for these planet-host stars.

Figure \ref{fig_param} shows the comparison between the new spectroscopic parameters 
derived in this work against the previous (in the literature) parameters reported in 
SWEET-Cat mostly coming from planet discovery papers. The blue circles represent new planet hosts for which, for the
first time, we use our method to derive parameters. The orange triangles represent the planet hosts for 
which we had already reported homogeneous parameters but using the small line list. 
As we already mentioned before we show a very good consistency between the new and  previous homogeneous parameters. The gain is mainly related to the use of 
more lines, which statistically are translated into smaller precision errors. For 
the completely new homogeneous data (blue circles; 106 planet hosts) we 
have a general good consistency with the literature data where the largest 
differences appear for evolved stars (lower surface gravities). The lower 
number of points for the microturbulence comes from the fact that many of the 
spectroscopic analyses available from the literature for these stars do 
not report this parameter.

\subsection{Different Instruments}

One of the main advantages of SWEET-Cat is the homogenization of the spectroscopic 
parameters derived by our team. A possible problem that could affect the 
homogenization would be the use of different instruments. Actually
the use of the same instrument, with the same configuration, is normally considered a 
very good argument to maintain the homogenization of the spectroscopic analysis. 
Moreover, there are works such as \citet[][]{Bedell-2014} stating 
that the largest effect is associated with the use of different instruments, which 
can go up to 0.04 dex in metallicity. Recently we had also shown that even using the 
same instrument, but with different individual observations of the 
same star at different times and conditions, can reveal some non-negligible 
differences in derived individual abundances \citep[][]{Adibekyan-2016}. Therefore, 
it is fundamental to understand how good the homogeneity is in SWEET-Cat when using 
different instruments.

\begin{figure*}
  \centering
  \includegraphics[width=19cm]{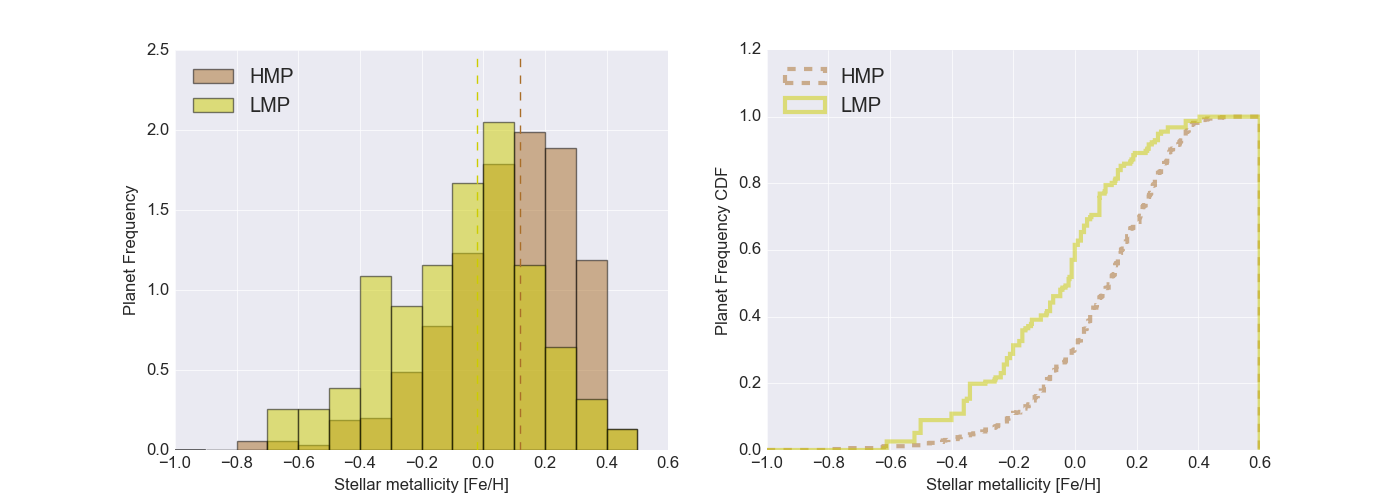}
  \includegraphics[width=19cm]{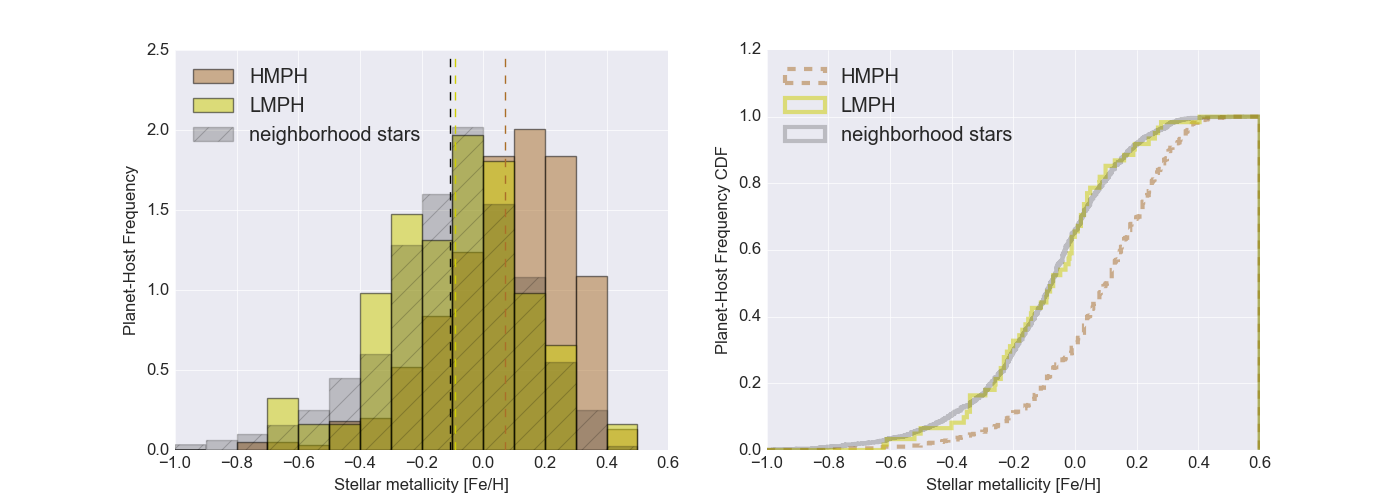}
  \caption{Metallicity distributions for the HMP and LMP samples (top panel - per planet), and HMPH, LMPH, and a sample representing neighborhood stars (bottom panel - per star). The dashed vertical lines represent the average of each distribution. The right panels shows the CDF 
  for a better comparison.}
  \label{planet_metallicity_hist}
\end{figure*}

To answer this relevant question we can make use of our already large compilation of 
spectroscopic data to test our method against the use of different 
instruments/configurations. Figure \ref{fig_inst} shows the differences found using ARES+MOOG on spectra of the same stars observed with different instruments. In this 
plot we merge different instruments and, although most of the comparisons are made 
with only two spectrographs, there is a wide variety of different pairings to access 
any possible biases with different instruments. Also there are a few stars that were 
observed by up to four to five different spectrographs. Looking at this figure and comparing 
the effective temperatures and [Fe/H] it is clear that the values derived are quite 
consistent when using different instruments. The standard deviation of the parameters 
relative to the averaged value, used as a reference for each star, is quite small at 
22 K for temperature and 0.01 dex for [Fe/H]. It is true that there are some large 
individual differences for temperatures that can be as high as 100 K in some cases, 
but they are actually well within the errors reported for those cases. Moreover, the 
average relative discrepancy from the average is consistent with or below one for all 
the instruments ($ < \frac{Teff_{inst} - <Teff>}{\sigma_{Teff_{inst}}} > \lesssim 
1$). This simple test demonstrates that our spectroscopic method is quite insensitive 
to the source of the spectroscopic data. It is worth mentioning that most of the 
spectroscopic data compiled is generally of very good quality, both in terms of 
S/N and resolution (see Figure \ref{fig_sn} and Table 
\ref{tab_spectrographs}).

In what regards the selection of the parameters to be listed in SWEET-Cat for the 
stars for which we have multiple instruments, we selected the values with the lowest 
estimated errors. We selected this instead of the average because there are 
cases in which the average can still be significantly affected by the presence of lower 
quality spectra from other instruments. This is of course strongly correlated 
with the number of lines used in each analysis and the individual precision of each 
line and therefore the spectroscopic data with higher resolution and higher S/N 
spectra are usually selected.

\section{Planet-host metallicity correlations}

One of the first observational constraints for planet formation theories was the 
correlation between the presence of giant planets and the higher metallicity of their 
host stars \citep[e.g.,][]{Santos-2004b, Fischer_Valenti-2005}. Of course, because of 
the known observational biases from the main planet detection methods, the first 
planets found were giants and massive. Very soon the correlation was tested for lower 
mass planets, where, despite the low statistic significance of the first results, the 
correlation was not observed \citep[e.g.,][]{Udry-2006,Sousa-2008, Ghezzi-2010, 
Sousa-2011, Buchhave-2012}. More recently, \citet[][]{Wang-2015} suggested that there 
should be an universal planet-metallicity correlation for all planets, but this might 
be related to the higher planet frequency and lower detectability of low-mass planets \citep[LMPs; see also][]{Zhu-2016}. Once settled, these observational correlations, 
should provide strong constrains for the theory of planet formation and evolution. In 
this section we take advantage of our largest sample of planet-host stars with 
uniform parameters to review the metallicity correlation.

Figure \ref{planet_metallicity_hist} shows the metallicity distribution of planet-host stars that have stellar parameters derived by our team in a homogeneous way. In 
the top panels we show the planet host counts by separating the population of high-mass planets (HMP; 700 planets, with minimum masses greater than 
30 M$_{\oplus}$) and LMPs (156 
planets). The location of the gap in the planet mass 
distribution presented in \citet{Mayor-2011} is 30 M$_{\oplus}$ and we also keep this threshold to be 
consistent with our previous works \citep[e.g.,][]{Sousa-2011b}. In this work, the metallicity of the host star appears multiple times for systems with multiple planets. On the bottom panels we plot the metallicity distribution of the planet-hosting stars. The planet with the highest minimum mass is considered for multiplanet systems. Three different 
population of stars are represented: \textit{i}) high-mass planet hosts (HMPH; 600 
stars, i.e., planet-host stars that have at least one orbiting planet with a minimum 
mass greater than 30 M$_{\oplus}$); \textit{ii}) low-mass planet hosts (LMPH; 61 stars); 
and \textit{iii}) a sample of 1111 stars located in the solar neighborhood
taken from the HARPS sample discussed in \citet[][]{Adibekyan-2012b}. We note that about 15\% of the solar neighborhood sample stars are known to host planets. Our goal is not to compare with nonhost stars, but to compare with the general distribution of stars.

Both cases (planet counts, and planet host counts) are represented in Fig. \ref{planet_metallicity_hist} where we show the metallicity distribution of the different populations in the left panels and in the right panels 
we present the cumulative distribution function (CDF) of the distributions that allow 
us to better compare the different distributions.

As expected, the increasing number of exoplanets continues to reinforce the 
metallicity correlation observed for massive planets, while for the LMP 
the distribution is clearly different, especially when we compare 
the LMPH with the neighborhood stars, where the resemblances are remarkable and using 
Kolmogorov-Smirnov (K-S) test we cannot clearly distinct these stars. Just for comparison, in 
\citet[][]{Sousa-2011b} we had only 10 planet-host stars with 
exo-Neptunes, while we are considering 61 planet-host stars. Table 
\ref{tab_stats_ks_ad} shows the statistics and p-values of the different populations 
comparisons that we discuss in this work.

Given the similarities between the metallicity distributions of LMPs and 
the neighborhood stars, and assuming that all these LMPs are included in 
stars in the solar neighborhood (which is not far from reality since
most of these planet-host stars are bright stars observed by RV follow up), this would 
of course lead to a flat distribution of the LMP frequency in the solar 
neighborhood as already shown in our previous works \citep[][]{Sousa-2008, 
Sousa-2011b}.

\begin{table}[t]
\caption[]{Kolmogorov-Smirnov test comparing the metallicities distribution.}
\small
  \begin{center}

    \begin{tabular}{lcc}
    \hline
    \hline
    [Fe/H] Samples &      K-S        &   K-S       \\
                   &   statistic     & p-value   \\
    \hline
    HMP vs. LMP    (1)  & 0.31 & 1.82e-11  \\
    HMPH vs. LMPH  (2)  & 0.37 & 2.49e-07  \\
    LMPH vs. stars (2)  & 0.06 & 9.68e-01  \\
    \hline
    \end{tabular}
  \end{center}
\label{tab_stats_ks_ad}
\tablefoot{(1) - Using planet counts; (2) - Using star counts. (see Fig. \ref{planet_metallicity_hist}); ''stars`` corresponds to the solar neighborhood sample.} 
\end{table}

\section{Summary}

In this work we update SWEET-Cat with new precise and homogeneous spectroscopic 
parameters using spectra that we collected from different sources. We describe 
in detail all the sources of our spectroscopic data used to compile this catalog. 

We also update some of the previous stellar parameters listed in the catalog where 
we confirm the consistency of our results. The new homogeneous parameters are also 
compared with literature values showing generally good agreement, but with a 
significant improvement for several specific host stars.
We demonstrate in this work that our spectroscopic analysis method to derive stellar  
parameters provides very consistent results when analyzing the same stars with many 
different high-resolution spectrographs.

Finally the known planet-host metallicity correlations are reviewed focusing on the 
differences found between stars hosting LMPs and stars hosting massive 
planets. While the giant planet hosts have the clear metallicity 
correlation, we show that the stars hosting LMPs (with a minimum mass 
below 30 M$_{\oplus}$) are indistinguishable from the solar neighborhood.

\longtab{
\begin{longtable}{lcccccccr}
\caption{\label{tab_parameters} Spectroscopic Parameters with ARES+MOOG}\\
\hline\hline
Star ID     & T$_{\mathrm{eff}}$ & $\log{g}_{spec}$ & $\xi_{\mathrm{t}}$ & \multicolumn{1}{c}{[Fe/H]} & N(\ion{Fe} {i},\ion{Fe}{ii}) & Mass      & Instrument  & SW \\
            &  [K]               & [cm\,s$^{-2}$]   &  [km\,s$^{-1}$]    &          [dex]             &                              & [$M_{\odot}$]&   &  \\

\hline
\endfirsthead
\caption{continued.}\\
\hline\hline
Star ID     & T$_{\mathrm{eff}}$ & $\log{g}_{spec}$ & $\xi_{\mathrm{t}}$ & \multicolumn{1}{c}{[Fe/H]} & N(\ion{Fe} {i},\ion{Fe}{ii}) & Mass      & Instrument  & SW \\
            &  [K]               & [cm\,s$^{-2}$]   &  [km\,s$^{-1}$]    &          [dex]             &                              & [$M_{\odot}$]&   &  \\

\hline
\endhead
\hline
\endfoot

\object{           BD-061339} & 4559 $\pm$\ 281 &  4.50 $\pm$\  0.69 &  0.57 $\pm$\  1.03 & -0.34 $\pm$\  0.08 &  94,  6 &  0.72 $\pm$\  0.21 &      HARPS & yes \\
\object{           BD+152940} & 4838 $\pm$\  46 &  2.72 $\pm$\  0.12 &  1.43 $\pm$\  0.04 & -0.15 $\pm$\  0.03 &  82,  9 &  2.07 $\pm$\  0.25 &       UVES & yes \\
\object{            BD+20594} & 5659 $\pm$\  18 &  4.28 $\pm$\  0.03 &  0.87 $\pm$\  0.03 & -0.14 $\pm$\  0.01 & 243, 31 &  0.95 $\pm$\  0.07 &      HARPS & yes \\
\object{            CoRoT-19} & 6425 $\pm$\  95 &  4.54 $\pm$\  0.09 &  1.52 $\pm$\  0.13 &  0.24 $\pm$\  0.07 & 215, 29 &  1.25 $\pm$\  0.09 &      HARPS & yes \\
\object{            CoRoT-25} & 6166 $\pm$\  38 &  4.41 $\pm$\  0.04 &  1.37 $\pm$\  0.05 &  0.16 $\pm$\  0.03 & 194, 23 &  1.16 $\pm$\  0.08 &       UVES & yes \\
\object{             CoRoT-6} & 6278 $\pm$\  73 &  4.76 $\pm$\  0.07 &  1.43 $\pm$\  0.14 & -0.05 $\pm$\  0.05 & 170, 22 &  1.05 $\pm$\  0.08 &       UVES & yes \\
\object{       EPIC201295312} & 5883 $\pm$\  29 &  4.20 $\pm$\  0.04 &  1.15 $\pm$\  0.03 &  0.20 $\pm$\  0.02 & 240, 31 &  1.17 $\pm$\  0.08 &      HARPS & yes \\
\object{               K2-19} & 5355 $\pm$\  35 &  4.46 $\pm$\  0.06 &  0.80 $\pm$\  0.06 &  0.05 $\pm$\  0.02 & 230, 33 &  0.86 $\pm$\  0.06 &      HARPS & yes \\
\object{               K2-24} & 5706 $\pm$\  33 &  4.36 $\pm$\  0.06 &  1.01 $\pm$\  0.04 &  0.41 $\pm$\  0.03 & 239, 33 &  1.09 $\pm$\  0.08 &      HARPS & yes \\
\object{               K2-32} & 5276 $\pm$\  37 &  4.38 $\pm$\  0.06 &  0.66 $\pm$\  0.07 & -0.05 $\pm$\  0.03 & 245, 34 &  0.84 $\pm$\  0.06 &      HARPS & yes \\
\object{              etaCet} & 4687 $\pm$\  85 &  2.63 $\pm$\  0.21 &  1.41 $\pm$\  0.08 &  0.15 $\pm$\  0.05 &  79, 10 &  2.25 $\pm$\  0.44 &       UVES & yes \\
\object{             GJ160.2} & 4498 $\pm$\ 152 &  4.35 $\pm$\  0.50 &  0.15 $\pm$\  1.40 & -0.26 $\pm$\  0.08 &  94,  6 &  0.70 $\pm$\  0.14 &      HARPS & yes \\
\object{            HAT-P-13} & 5797 $\pm$\  34 &  4.14 $\pm$\  0.06 &  1.03 $\pm$\  0.04 &  0.44 $\pm$\  0.03 & 231, 32 &  1.27 $\pm$\  0.09 &     SOPHIE & yes \\
\object{            HAT-P-45} & 6394 $\pm$\  94 &  4.58 $\pm$\  0.08 &  1.38 $\pm$\  0.14 &  0.01 $\pm$\  0.07 & 169, 21 &  1.14 $\pm$\  0.08 &       UVES & yes \\
\object{            HAT-P-50} & 6283 $\pm$\  82 &  4.17 $\pm$\  0.11 &  1.63 $\pm$\  0.13 & -0.11 $\pm$\  0.06 & 166, 21 &  1.23 $\pm$\  0.10 &       UVES & yes \\
\object{            HAT-P-54} & 4505 $\pm$\ 163 &  4.22 $\pm$\  0.56 &  0.91 $\pm$\  0.38 & -0.04 $\pm$\  0.06 &  83,  8 &  0.75 $\pm$\  0.20 &       UVES & yes \\
\object{             HATS-13} & 5526 $\pm$\  71 &  4.62 $\pm$\  0.12 &  1.00 $\pm$\  0.12 &  0.10 $\pm$\  0.05 & 228, 34 &  0.89 $\pm$\  0.07 &      FEROS & yes \\
\object{              HATS-2} & 5233 $\pm$\  99 &  4.45 $\pm$\  0.19 &  1.62 $\pm$\  0.12 &  0.22 $\pm$\  0.05 & 213, 32 &  0.88 $\pm$\  0.08 &      FEROS & yes \\
\object{            HD113337} & 6918 $\pm$\  47 &  4.71 $\pm$\  0.07 &  1.82 $\pm$\  0.07 &  0.25 $\pm$\  0.03 & 214, 34 &  1.38 $\pm$\  0.09 &     SOPHIE & yes \\
\object{            HD128356} & 4932 $\pm$\ 126 &  4.31 $\pm$\  0.29 &  0.64 $\pm$\  0.25 &  0.25 $\pm$\  0.06 & 109, 14 &  0.84 $\pm$\  0.11 &      HARPS & yes \\
\object{             HD14067} & 4883 $\pm$\  49 &  2.76 $\pm$\  0.11 &  1.52 $\pm$\  0.05 & -0.08 $\pm$\  0.04 &  87, 11 &  2.07 $\pm$\  0.23 &       UVES & yes \\
\object{            HD141399} & 5602 $\pm$\  34 &  4.24 $\pm$\  0.05 &  0.90 $\pm$\  0.05 &  0.36 $\pm$\  0.03 & 233, 34 &  1.09 $\pm$\  0.08 &     SOPHIE & yes \\
\object{            HD142245} & 4843 $\pm$\  66 &  3.21 $\pm$\  0.14 &  1.18 $\pm$\  0.07 &  0.13 $\pm$\  0.04 &  81, 10 &  1.53 $\pm$\  0.20 &       UVES & yes \\
\object{            HD143761} & 5829 $\pm$\  14 &  4.29 $\pm$\  0.02 &  1.02 $\pm$\  0.02 & -0.20 $\pm$\  0.01 & 242, 30 &  0.98 $\pm$\  0.07 &     SOPHIE & yes \\
\object{            HD147873} & 6191 $\pm$\  30 &  4.14 $\pm$\  0.04 &  1.71 $\pm$\  0.04 &  0.24 $\pm$\  0.02 & 237, 32 &  1.35 $\pm$\  0.09 &      HARPS & yes \\
\object{            HD155233} & 4842 $\pm$\  45 &  3.16 $\pm$\  0.12 &  1.16 $\pm$\  0.04 &  0.06 $\pm$\  0.03 & 112, 15 &  1.57 $\pm$\  0.18 &      FEROS & yes \\
\object{            HD156668} & 4804 $\pm$\ 141 &  4.44 $\pm$\  0.33 &  0.42 $\pm$\  0.57 & -0.02 $\pm$\  0.06 &  90,  9 &  0.75 $\pm$\  0.10 &       UVES & yes \\
\object{            HD162004} & 6273 $\pm$\  33 &  4.50 $\pm$\  0.04 &  1.23 $\pm$\  0.04 &  0.08 $\pm$\  0.02 & 219, 30 &  1.14 $\pm$\  0.08 &     SOPHIE & yes \\
\object{            HD164595} & 5725 $\pm$\  14 &  4.41 $\pm$\  0.03 &  0.90 $\pm$\  0.03 & -0.09 $\pm$\  0.01 & 242, 32 &  0.94 $\pm$\  0.07 &     SOPHIE & yes \\
\object{            HD165155} & 5387 $\pm$\  31 &  4.39 $\pm$\  0.05 &  0.81 $\pm$\  0.05 &  0.12 $\pm$\  0.02 & 237, 33 &  0.90 $\pm$\  0.07 &      HARPS & yes \\
\object{              HD1666} & 6510 $\pm$\  55 &  4.25 $\pm$\  0.06 &  1.77 $\pm$\  0.07 &  0.39 $\pm$\  0.04 & 224, 29 &  1.44 $\pm$\  0.10 &      HARPS & yes \\
\object{              HD1666} & 6497 $\pm$\  35 &  4.20 $\pm$\  0.06 &  1.67 $\pm$\  0.04 &  0.38 $\pm$\  0.03 & 187, 22 &  1.46 $\pm$\  0.10 &       UVES &  no \\
\object{            HD189733} & 5080 $\pm$\  91 &  4.51 $\pm$\  0.20 &  0.74 $\pm$\  0.17 & -0.02 $\pm$\  0.04 &  95, 13 &  0.79 $\pm$\  0.07 &    CORALIE &  no \\
\object{            HD189733} & 4984 $\pm$\  59 &  4.40 $\pm$\  0.13 &  0.68 $\pm$\  0.14 & -0.04 $\pm$\  0.03 & 115, 13 &  0.78 $\pm$\  0.06 &   ESPADONS &  no \\
\object{            HD189733} & 5010 $\pm$\  48 &  4.44 $\pm$\  0.12 &  0.83 $\pm$\  0.10 & -0.05 $\pm$\  0.02 & 112, 12 &  0.77 $\pm$\  0.06 &      FEROS &  no \\
\object{            HD189733} & 4969 $\pm$\  43 &  4.31 $\pm$\  0.10 &  0.76 $\pm$\  0.10 & -0.07 $\pm$\  0.02 & 114, 14 &  0.78 $\pm$\  0.06 &      HARPS & yes \\
\object{            HD222076} & 4834 $\pm$\  59 &  3.24 $\pm$\  0.13 &  1.15 $\pm$\  0.06 &  0.16 $\pm$\  0.03 & 112, 13 &  1.50 $\pm$\  0.18 &      FEROS & yes \\
\object{            HD224538} & 6235 $\pm$\  24 &  4.37 $\pm$\  0.04 &  1.34 $\pm$\  0.03 &  0.37 $\pm$\  0.02 & 247, 34 &  1.30 $\pm$\  0.09 &      HARPS & yes \\
\object{             HD32963} & 5789 $\pm$\  33 &  4.46 $\pm$\  0.05 &  1.01 $\pm$\  0.05 &  0.12 $\pm$\  0.03 & 231, 32 &  1.00 $\pm$\  0.07 &      FEROS &  no \\
\object{             HD32963} & 5767 $\pm$\  23 &  4.38 $\pm$\  0.03 &  0.83 $\pm$\  0.04 &  0.09 $\pm$\  0.02 & 235, 32 &  1.01 $\pm$\  0.07 &     SOPHIE & yes \\
\object{             HD33844} & 4832 $\pm$\  63 &  3.01 $\pm$\  0.14 &  1.21 $\pm$\  0.06 &  0.18 $\pm$\  0.04 & 110, 14 &  1.79 $\pm$\  0.24 &      FEROS & yes \\
\object{             HD42618} & 5751 $\pm$\  11 &  4.45 $\pm$\  0.01 &  0.94 $\pm$\  0.02 & -0.09 $\pm$\  0.01 & 246, 29 &  0.94 $\pm$\  0.07 &      HARPS & yes \\
\object{             HD42618} & 5736 $\pm$\  16 &  4.44 $\pm$\  0.03 &  0.85 $\pm$\  0.03 & -0.11 $\pm$\  0.01 & 243, 31 &  0.93 $\pm$\  0.07 &     SOPHIE &  no \\
\object{             HD47366} & 4919 $\pm$\  37 &  3.08 $\pm$\  0.08 &  1.26 $\pm$\  0.04 & -0.00 $\pm$\  0.03 & 114, 14 &  1.67 $\pm$\  0.15 &      HARPS & yes \\
\object{             HD47366} & 4914 $\pm$\  42 &  3.06 $\pm$\  0.09 &  1.24 $\pm$\  0.04 &  0.01 $\pm$\  0.03 &  89, 11 &  1.69 $\pm$\  0.16 &       UVES &  no \\
\object{              HD4747} & 5333 $\pm$\  23 &  4.45 $\pm$\  0.06 &  0.76 $\pm$\  0.05 & -0.21 $\pm$\  0.02 & 241, 30 &  0.81 $\pm$\  0.06 &      FEROS & yes \\
\object{             HD68402} & 5907 $\pm$\  33 &  4.43 $\pm$\  0.04 &  1.04 $\pm$\  0.04 &  0.27 $\pm$\  0.03 & 238, 31 &  1.10 $\pm$\  0.08 &      HARPS & yes \\
\object{             HD72892} & 5685 $\pm$\  29 &  4.33 $\pm$\  0.04 &  0.95 $\pm$\  0.04 &  0.15 $\pm$\  0.02 & 239, 33 &  1.02 $\pm$\  0.07 &      HARPS & yes \\
\object{              HD9174} & 5631 $\pm$\  30 &  4.05 $\pm$\  0.04 &  1.12 $\pm$\  0.03 &  0.36 $\pm$\  0.02 & 243, 33 &  1.22 $\pm$\  0.08 &      HARPS & yes \\
\object{             HD95872} & 5304 $\pm$\  71 &  4.41 $\pm$\  0.13 &  0.92 $\pm$\  0.10 &  0.26 $\pm$\  0.04 & 172, 24 &  0.91 $\pm$\  0.08 &       UVES & yes \\
\object{           HIP105854} & 4575 $\pm$\ 102 &  2.43 $\pm$\  0.27 &  1.47 $\pm$\  0.09 &  0.21 $\pm$\  0.05 & 102, 14 &  2.56 $\pm$\  0.63 &      FEROS &  no \\
\object{           HIP105854} & 4618 $\pm$\ 113 &  2.36 $\pm$\  0.28 &  1.47 $\pm$\  0.11 &  0.17 $\pm$\  0.07 &  74, 10 &  2.74 $\pm$\  0.73 &       UVES & yes \\
\object{            HIP63242} & 4812 $\pm$\  37 &  2.51 $\pm$\  0.09 &  1.68 $\pm$\  0.04 & -0.31 $\pm$\  0.03 & 113, 12 &  2.32 $\pm$\  0.23 &      FEROS & yes \\
\object{            HIP65891} & 4928 $\pm$\  43 &  2.89 $\pm$\  0.12 &  1.38 $\pm$\  0.04 &  0.12 $\pm$\  0.03 & 108, 14 &  1.98 $\pm$\  0.24 &      FEROS & yes \\
\object{            HIP67537} & 4976 $\pm$\  46 &  2.95 $\pm$\  0.11 &  1.40 $\pm$\  0.04 &  0.17 $\pm$\  0.03 & 109, 14 &  1.95 $\pm$\  0.22 &      FEROS & yes \\
\object{            HIP67851} & 4805 $\pm$\  39 &  3.19 $\pm$\  0.14 &  1.13 $\pm$\  0.05 &  0.01 $\pm$\  0.03 & 112, 14 &  1.49 $\pm$\  0.19 &      FEROS &  no \\
\object{            HIP67851} & 4809 $\pm$\  51 &  3.09 $\pm$\  0.12 &  1.11 $\pm$\  0.06 & -0.01 $\pm$\  0.03 &  88, 11 &  1.60 $\pm$\  0.18 &       UVES & yes \\
\object{            HIP68468} & 5840 $\pm$\  12 &  4.34 $\pm$\  0.01 &  1.08 $\pm$\  0.02 &  0.08 $\pm$\  0.01 & 240, 32 &  1.04 $\pm$\  0.07 &      HARPS & yes \\
\object{            HIP74890} & 4808 $\pm$\  69 &  3.09 $\pm$\  0.17 &  1.24 $\pm$\  0.07 &  0.21 $\pm$\  0.04 & 109, 15 &  1.68 $\pm$\  0.25 &      FEROS &  no \\
\object{            HIP74890} & 4848 $\pm$\  84 &  3.33 $\pm$\  0.20 &  1.28 $\pm$\  0.10 &  0.20 $\pm$\  0.05 &  85, 11 &  1.45 $\pm$\  0.23 &       UVES & yes \\
\object{             HIP8541} & 4616 $\pm$\  46 &  2.89 $\pm$\  0.13 &  1.18 $\pm$\  0.04 & -0.08 $\pm$\  0.02 & 112, 14 &  1.71 $\pm$\  0.21 &      FEROS & yes \\
\object{            HIP97233} & 4955 $\pm$\  60 &  3.24 $\pm$\  0.14 &  1.24 $\pm$\  0.06 &  0.27 $\pm$\  0.04 & 111, 14 &  1.61 $\pm$\  0.20 &      FEROS & yes \\
\object{               K2-31} & 5406 $\pm$\  33 &  4.43 $\pm$\  0.06 &  0.95 $\pm$\  0.07 &  0.16 $\pm$\  0.02 & 235, 31 &  0.91 $\pm$\  0.07 &      HARPS & yes \\
\object{             KELT-10} & 5850 $\pm$\  37 &  4.40 $\pm$\  0.04 &  1.05 $\pm$\  0.05 &  0.13 $\pm$\  0.03 & 196, 21 &  1.04 $\pm$\  0.07 &       UVES & yes \\
\object{             KELT-15} & 6428 $\pm$\  72 &  4.58 $\pm$\  0.08 &  1.77 $\pm$\  0.09 &  0.24 $\pm$\  0.05 & 162, 20 &  1.24 $\pm$\  0.09 &       UVES & yes \\
\object{              KELT-8} & 5804 $\pm$\  37 &  4.32 $\pm$\  0.06 &  1.16 $\pm$\  0.05 &  0.25 $\pm$\  0.03 & 198, 24 &  1.09 $\pm$\  0.08 &       UVES & yes \\
\object{           Kepler-10} & 5685 $\pm$\  27 &  4.35 $\pm$\  0.04 &  0.64 $\pm$\  0.05 & -0.14 $\pm$\  0.02 & 240, 29 &  0.93 $\pm$\  0.07 &     SOPHIE & yes \\
\object{           Kepler-68} & 5884 $\pm$\  45 &  4.35 $\pm$\  0.09 &  0.99 $\pm$\  0.06 &  0.15 $\pm$\  0.04 & 227, 32 &  1.08 $\pm$\  0.08 &     SOPHIE & yes \\
\object{           Kepler-93} & 5624 $\pm$\  40 &  4.48 $\pm$\  0.08 &  0.72 $\pm$\  0.08 & -0.15 $\pm$\  0.03 & 226, 28 &  0.89 $\pm$\  0.07 &     SOPHIE & yes \\
\object{              Pr0201} & 6247 $\pm$\  46 &  4.51 $\pm$\  0.06 &  1.40 $\pm$\  0.06 &  0.25 $\pm$\  0.03 & 187, 24 &  1.19 $\pm$\  0.08 &       UVES & yes \\
\object{            WASP-101} & 6503 $\pm$\ 100 &  4.68 $\pm$\  0.11 &  1.81 $\pm$\  0.17 &  0.18 $\pm$\  0.07 & 210, 31 &  1.22 $\pm$\  0.09 &      HARPS & yes \\
\object{            WASP-101} & 6534 $\pm$\ 132 &  4.89 $\pm$\  0.11 &  1.78 $\pm$\  0.23 &  0.18 $\pm$\  0.09 & 152, 19 &  1.23 $\pm$\  0.09 &       UVES &  no \\
\object{            WASP-104} & 5416 $\pm$\  86 &  4.36 $\pm$\  0.16 &  0.76 $\pm$\  0.16 &  0.40 $\pm$\  0.06 & 187, 25 &  0.99 $\pm$\  0.09 &       UVES & yes \\
\object{            WASP-105} & 5183 $\pm$\  66 &  4.23 $\pm$\  0.16 &  0.72 $\pm$\  0.14 &  0.36 $\pm$\  0.04 & 111, 15 &  0.95 $\pm$\  0.09 &      HARPS & yes \\
\object{            WASP-106} & 6265 $\pm$\  36 &  4.38 $\pm$\  0.04 &  1.39 $\pm$\  0.04 &  0.15 $\pm$\  0.03 & 225, 32 &  1.21 $\pm$\  0.08 &      HARPS & yes \\
\object{            WASP-108} & 6193 $\pm$\  33 &  4.47 $\pm$\  0.04 &  1.27 $\pm$\  0.04 &  0.26 $\pm$\  0.02 & 189, 23 &  1.18 $\pm$\  0.08 &       UVES & yes \\
\object{            WASP-111} & 6698 $\pm$\ 125 &  4.78 $\pm$\  0.09 &  2.42 $\pm$\  0.21 &  0.25 $\pm$\  0.08 & 155, 20 &  1.32 $\pm$\  0.10 &       UVES & yes \\
\object{            WASP-117} & 6026 $\pm$\  17 &  4.35 $\pm$\  0.02 &  1.17 $\pm$\  0.03 & -0.13 $\pm$\  0.01 & 245, 32 &  1.04 $\pm$\  0.07 &      HARPS & yes \\
\object{            WASP-122} & 5859 $\pm$\  41 &  4.31 $\pm$\  0.05 &  1.18 $\pm$\  0.05 &  0.37 $\pm$\  0.03 & 189, 21 &  1.16 $\pm$\  0.08 &       UVES & yes \\
\object{            WASP-126} & 5807 $\pm$\  39 &  4.39 $\pm$\  0.06 &  1.09 $\pm$\  0.05 &  0.19 $\pm$\  0.03 & 185, 23 &  1.05 $\pm$\  0.08 &       UVES & yes \\
\object{            WASP-129} & 5983 $\pm$\  23 &  4.38 $\pm$\  0.04 &  1.13 $\pm$\  0.03 &  0.16 $\pm$\  0.02 & 234, 32 &  1.11 $\pm$\  0.08 &      HARPS & yes \\
\object{            WASP-130} & 5667 $\pm$\  34 &  4.43 $\pm$\  0.05 &  0.96 $\pm$\  0.05 &  0.31 $\pm$\  0.03 & 247, 34 &  1.02 $\pm$\  0.07 &      HARPS & yes \\
\object{            WASP-131} & 6143 $\pm$\  23 &  4.18 $\pm$\  0.04 &  1.33 $\pm$\  0.03 & -0.01 $\pm$\  0.02 & 245, 34 &  1.20 $\pm$\  0.08 &      HARPS & yes \\
\object{            WASP-132} & 4742 $\pm$\ 201 &  4.23 $\pm$\  0.50 &  0.40 $\pm$\  0.86 &  0.21 $\pm$\  0.09 & 106, 13 &  0.84 $\pm$\  0.21 &      HARPS & yes \\
\object{            WASP-139} & 5109 $\pm$\  70 &  4.40 $\pm$\  0.15 &  0.60 $\pm$\  0.16 &  0.05 $\pm$\  0.04 & 111, 14 &  0.82 $\pm$\  0.07 &      HARPS & yes \\
\object{             WASP-20} & 5987 $\pm$\  20 &  4.33 $\pm$\  0.03 &  1.19 $\pm$\  0.03 &  0.07 $\pm$\  0.02 & 247, 35 &  1.09 $\pm$\  0.08 &      HARPS & yes \\
\object{             WASP-39} & 5512 $\pm$\  40 &  4.36 $\pm$\  0.06 &  0.72 $\pm$\  0.07 &  0.04 $\pm$\  0.03 & 184, 24 &  0.93 $\pm$\  0.07 &       UVES & yes \\
\object{             WASP-43} & 4798 $\pm$\ 216 &  4.55 $\pm$\  0.71 &  0.98 $\pm$\  0.52 & -0.13 $\pm$\  0.08 &  65,  6 &  0.80 $\pm$\  0.24 &       UVES & yes \\
\object{             WASP-46} & 5725 $\pm$\  39 &  4.47 $\pm$\  0.06 &  0.82 $\pm$\  0.07 & -0.18 $\pm$\  0.03 & 191, 23 &  0.91 $\pm$\  0.07 &       UVES & yes \\
\object{             WASP-48} & 6334 $\pm$\ 103 &  4.55 $\pm$\  0.15 &  1.86 $\pm$\  0.17 & -0.02 $\pm$\  0.07 & 194, 25 &  1.12 $\pm$\  0.09 &     SOPHIE & yes \\
\object{             WASP-49} & 5540 $\pm$\  18 &  4.36 $\pm$\  0.03 &  0.82 $\pm$\  0.03 & -0.07 $\pm$\  0.01 & 247, 31 &  0.91 $\pm$\  0.06 &      HARPS & yes \\
\object{             WASP-49} & 5534 $\pm$\  28 &  4.31 $\pm$\  0.04 &  0.80 $\pm$\  0.05 & -0.08 $\pm$\  0.02 & 195, 23 &  0.92 $\pm$\  0.07 &       UVES &  no \\
\object{             WASP-68} & 5985 $\pm$\  21 &  4.19 $\pm$\  0.03 &  1.29 $\pm$\  0.02 &  0.34 $\pm$\  0.02 & 236, 32 &  1.28 $\pm$\  0.08 &      HARPS & yes \\
\object{             WASP-69} & 4765 $\pm$\ 120 &  4.13 $\pm$\  0.31 &  0.79 $\pm$\  0.30 &  0.30 $\pm$\  0.06 & 105, 15 &  0.86 $\pm$\  0.14 &      HARPS & yes \\
\object{            WASP-70A} & 5864 $\pm$\  25 &  4.36 $\pm$\  0.03 &  1.01 $\pm$\  0.03 &  0.21 $\pm$\  0.02 & 191, 24 &  1.09 $\pm$\  0.07 &       UVES & yes \\
\object{             WASP-74} & 6075 $\pm$\  43 &  4.43 $\pm$\  0.06 &  1.26 $\pm$\  0.05 &  0.48 $\pm$\  0.03 & 200, 23 &  1.24 $\pm$\  0.08 &       UVES & yes \\
\object{             WASP-90} & 6339 $\pm$\  65 &  4.28 $\pm$\  0.04 &  1.58 $\pm$\  0.08 &  0.14 $\pm$\  0.05 & 168, 23 &  1.29 $\pm$\  0.09 &       UVES & yes \\
\object{                XO-4} & 6707 $\pm$\  86 &  4.79 $\pm$\  0.07 &  1.87 $\pm$\  0.14 &  0.02 $\pm$\  0.05 & 198, 31 &  1.22 $\pm$\  0.09 &     SOPHIE & yes \\
\object{             YBP1194} & 5970 $\pm$\  26 &  4.42 $\pm$\  0.05 &  1.13 $\pm$\  0.03 & -0.00 $\pm$\  0.02 & 192, 23 &  1.03 $\pm$\  0.07 &       UVES & yes \\
\object{             YBP1514} & 5076 $\pm$\  34 &  3.61 $\pm$\  0.09 &  0.98 $\pm$\  0.04 &  0.09 $\pm$\  0.02 &  91, 10 &  1.23 $\pm$\  0.11 &       UVES & yes \\
\object{              YBP401} & 6131 $\pm$\  37 &  4.36 $\pm$\  0.04 &  1.13 $\pm$\  0.05 &  0.04 $\pm$\  0.03 & 238, 33 &  1.13 $\pm$\  0.08 &      HARPS & yes \\
\end{longtable}
}

\longtab{
\begin{longtable}{lccccccccr}
\caption{\label{tab_parameters_syn} Spectroscopic Parameters with synthesis}\\
\hline\hline
Star ID     & T$_{\mathrm{eff}}$ & $\log{g}_{spec}$ &  $v_{\mathrm{mic}}$ & \multicolumn{1}{c}{[Fe/H]} &  $v \sin(i)$         & $v_{\mathrm{mac}}$ & Mass         & Instrument  & SW \\
            &  [K]               & [cm\,s$^{-2}$]   &  [km\,s$^{-1}$]     &         [dex]              &  [km\,s$^{-1}$]      & [km\,s$^{-1}$]     & [$M_{\odot}$]&             &  \\
\hline
\endfirsthead
\caption{continued.}\\
\hline\hline
Star ID     & T$_{\mathrm{eff}}$ & $\log{g}_{spec}$ &  $v_{\mathrm{mic}}$ & \multicolumn{1}{c}{[Fe/H]} &  $v \sin(i)$         & $v_{\mathrm{mac}}$ & Mass         & Instrument  & SW \\
            &  [K]               & [cm\,s$^{-2}$]   &  [km\,s$^{-1}$]     &         [dex]              &  [km\,s$^{-1}$]      & [km\,s$^{-1}$]     & [$M_{\odot}$]&             &  \\

\hline
\endhead
\hline
\endfoot
\object{               38Vir} & 6440 $\pm$\  51 &  4.42 $\pm$\  0.22 &  1.44 &  0.16 $\pm$\  0.05 & 28.10 $\pm$\  1.30 &  4.92 &  1.29 $\pm$\  0.12 &      HARPS & yes \\
\object{       EPIC211990866} & 5981 $\pm$\  44 &  4.32 $\pm$\  0.15 &  1.19 &  0.18 $\pm$\  0.04 & 13.60 $\pm$\  0.30 &  4.19 &  1.15 $\pm$\  0.10 &       UVES & yes \\
\object{            HAT-P-56} & 6491 $\pm$\  72 &  4.26 $\pm$\  0.30 &  1.73 & -0.22 $\pm$\  0.03 & 33.60 $\pm$\  1.30 &  4.82 &  1.25 $\pm$\  0.16 &       UVES & yes \\
\object{            WASP-103} & 6013 $\pm$\  44 &  4.24 $\pm$\  0.15 &  1.17 &  0.08 $\pm$\  0.04 & 10.10 $\pm$\  0.30 &  3.82 &  1.16 $\pm$\  0.10 &       FIES & yes \\
\object{            WASP-109} & 6573 $\pm$\  44 &  3.94 $\pm$\  0.15 &  1.76 & -0.10 $\pm$\  0.04 & 15.70 $\pm$\  0.30 &  5.20 &  1.49 $\pm$\  0.15 &       UVES & yes \\
\object{            WASP-120} & 6657 $\pm$\  44 &  4.41 $\pm$\  0.15 &  1.63 &  0.15 $\pm$\  0.04 & 15.00 $\pm$\  0.30 &  5.27 &  1.35 $\pm$\  0.10 &       UVES & yes \\
\object{            WASP-87A} & 6638 $\pm$\  44 &  4.06 $\pm$\  0.15 &  1.73 & -0.21 $\pm$\  0.04 & 10.40 $\pm$\  0.30 &  4.82 &  1.39 $\pm$\  0.13 &       UVES & yes \\
\object{           Aldebaran} & 3999 $\pm$\  25 &  1.67 $\pm$\  0.09 &  2.24 & -0.24 $\pm$\  0.03 &  2.50 $\pm$\  0.60 &  6.05 &  3.28 $\pm$\  0.31 &       UVES & yes \\
\object{             betaCnc} & 4077 $\pm$\  25 &  1.34 $\pm$\  0.09 &  2.04 & -0.30 $\pm$\  0.03 &  0.20 $\pm$\  0.10 &  6.32 &  4.19 $\pm$\  0.37 &       UVES & yes \\
\object{            HIP70849} & 4103 $\pm$\  25 &  3.70 $\pm$\  0.09 &  0.10 &  0.00 $\pm$\  0.03 &  0.30 $\pm$\  0.30 &  6.17 &  0.76 $\pm$\  0.07 &      HARPS & yes \\
\object{            HD208527} & 4118 $\pm$\  25 &  1.83 $\pm$\  0.09 &  2.44 &  0.15 $\pm$\  0.03 &  5.90 $\pm$\  0.60 &  5.98 &  3.37 $\pm$\  0.33 &       UVES & yes \\

\end{longtable}
}

\longtab{
\begin{longtable}{lcccccccr}
\caption{\label{tab_parameters_old} Spectroscopic Parameters with ARES+MOOG for old data.}\\
\hline\hline
Star ID     & T$_{\mathrm{eff}}$ & $\log{g}_{spec}$ & $\xi_{\mathrm{t}}$ & \multicolumn{1}{c}{[Fe/H]} & N(\ion{Fe} {i},\ion{Fe}{ii}) & Mass      & Instrument  & SW \\
            &  [K]               & [cm\,s$^{-2}$]   &  [km\,s$^{-1}$]    &          [dex]             &                              & [$M_{\odot}$]&   &  \\

\hline
\endfirsthead
\caption{continued.}\\
\hline\hline
Star ID     & T$_{\mathrm{eff}}$ & $\log{g}_{spec}$ & $\xi_{\mathrm{t}}$ & \multicolumn{1}{c}{[Fe/H]} & N(\ion{Fe} {i},\ion{Fe}{ii}) & Mass      & Instrument  & SW \\
            &  [K]               & [cm\,s$^{-2}$]   &  [km\,s$^{-1}$]    &          [dex]             &                              & [$M_{\odot}$]&   &  \\

\hline
\endhead
\hline
\endfoot

\object{               14Her} & 5452 $\pm$\  55 &  4.31 $\pm$\  0.16 &  1.10 $\pm$\  0.10 &  0.39 $\pm$\  0.04 & 196, 25 &  1.02 $\pm$\  0.09 &     ELODIE &  no \\
\object{               14Her} & 5368 $\pm$\  94 &  4.33 $\pm$\  0.15 &  0.92 $\pm$\  0.14 &  0.44 $\pm$\  0.06 & 214, 27 &  1.00 $\pm$\  0.09 &       SARG &  no \\
\object{               14Her} & 5286 $\pm$\  58 &  4.24 $\pm$\  0.11 &  0.80 $\pm$\  0.09 &  0.38 $\pm$\  0.04 & 222, 35 &  0.98 $\pm$\  0.08 &     SOPHIE & yes \\
\object{              16CygB} & 5827 $\pm$\  39 &  4.30 $\pm$\  0.06 &  0.97 $\pm$\  0.05 &  0.14 $\pm$\  0.03 & 217, 27 &  1.07 $\pm$\  0.08 &     ELODIE &  no \\
\object{              16CygB} & 5783 $\pm$\  19 &  4.42 $\pm$\  0.03 &  0.96 $\pm$\  0.03 &  0.09 $\pm$\  0.01 & 248, 33 &  1.00 $\pm$\  0.07 &   ESPADONS & yes \\
\object{              16CygB} & 5785 $\pm$\  28 &  4.41 $\pm$\  0.04 &  1.02 $\pm$\  0.04 &  0.10 $\pm$\  0.02 & 226, 28 &  1.01 $\pm$\  0.07 &       SARG &  no \\
\object{               51Peg} & 5854 $\pm$\  24 &  4.36 $\pm$\  0.04 &  1.10 $\pm$\  0.03 &  0.26 $\pm$\  0.02 & 229, 26 &  1.10 $\pm$\  0.08 &      FEROS &  no \\
\object{               51Peg} & 5814 $\pm$\  19 &  4.35 $\pm$\  0.03 &  1.04 $\pm$\  0.03 &  0.21 $\pm$\  0.01 & 197, 24 &  1.07 $\pm$\  0.07 &       UVES & yes \\
\object{               55Cnc} & 5346 $\pm$\  80 &  4.26 $\pm$\  0.18 &  1.29 $\pm$\  0.10 &  0.25 $\pm$\  0.05 & 207, 23 &  0.97 $\pm$\  0.10 &     ELODIE &  no \\
\object{               55Cnc} & 5353 $\pm$\  62 &  4.30 $\pm$\  0.14 &  1.01 $\pm$\  0.10 &  0.30 $\pm$\  0.04 & 183, 25 &  0.97 $\pm$\  0.08 &       UVES & yes \\
\object{               61Vir} & 5569 $\pm$\  24 &  4.39 $\pm$\  0.04 &  0.84 $\pm$\  0.04 &  0.03 $\pm$\  0.02 & 249, 33 &  0.93 $\pm$\  0.07 &      FEROS &  no \\
\object{               61Vir} & 5559 $\pm$\  17 &  4.36 $\pm$\  0.03 &  0.80 $\pm$\  0.03 & -0.01 $\pm$\  0.01 & 244, 31 &  0.93 $\pm$\  0.07 &      HARPS & yes \\
\object{               61Vir} & 5580 $\pm$\  21 &  4.39 $\pm$\  0.03 &  0.82 $\pm$\  0.04 &  0.02 $\pm$\  0.02 & 242, 28 &  0.93 $\pm$\  0.07 &     NARVAL &  no \\
\object{               61Vir} & 5556 $\pm$\  15 &  4.35 $\pm$\  0.03 &  0.84 $\pm$\  0.03 & -0.02 $\pm$\  0.01 & 197, 22 &  0.93 $\pm$\  0.07 &       UVES &  no \\
\object{               70Vir} & 5568 $\pm$\  29 &  4.00 $\pm$\  0.06 &  1.00 $\pm$\  0.04 & -0.02 $\pm$\  0.02 & 213, 22 &  1.09 $\pm$\  0.08 &     ELODIE &  no \\
\object{               70Vir} & 5565 $\pm$\  25 &  4.05 $\pm$\  0.05 &  1.05 $\pm$\  0.03 & -0.03 $\pm$\  0.02 & 234, 27 &  1.05 $\pm$\  0.08 &       SARG & yes \\
\object{           BD-103166} & 5367 $\pm$\  56 &  4.36 $\pm$\  0.12 &  1.02 $\pm$\  0.08 &  0.30 $\pm$\  0.04 & 222, 30 &  0.95 $\pm$\  0.08 &      FEROS & yes \\
\object{              GJ3021} & 5492 $\pm$\  37 &  4.45 $\pm$\  0.06 &  1.16 $\pm$\  0.05 &  0.11 $\pm$\  0.03 & 221, 29 &  0.91 $\pm$\  0.07 &    CORALIE &  no \\
\object{              GJ3021} & 5557 $\pm$\  36 &  4.47 $\pm$\  0.07 &  1.05 $\pm$\  0.06 &  0.13 $\pm$\  0.03 & 237, 32 &  0.93 $\pm$\  0.07 &      HARPS & yes \\
\object{            HD102195} & 5293 $\pm$\  34 &  4.37 $\pm$\  0.07 &  0.93 $\pm$\  0.07 &  0.03 $\pm$\  0.02 & 229, 30 &  0.86 $\pm$\  0.06 &   ESPADONS &  no \\
\object{            HD102195} & 5311 $\pm$\  29 &  4.44 $\pm$\  0.06 &  0.85 $\pm$\  0.05 &  0.04 $\pm$\  0.02 & 228, 32 &  0.85 $\pm$\  0.06 &      HARPS & yes \\
\object{            HD106252} & 5887 $\pm$\  25 &  4.41 $\pm$\  0.06 &  1.09 $\pm$\  0.04 & -0.04 $\pm$\  0.02 & 237, 24 &  1.00 $\pm$\  0.07 &      FEROS &  no \\
\object{            HD106252} & 5871 $\pm$\  15 &  4.38 $\pm$\  0.03 &  1.10 $\pm$\  0.02 & -0.07 $\pm$\  0.01 & 200, 23 &  1.00 $\pm$\  0.07 &       UVES & yes \\
\object{             HD10697} & 5711 $\pm$\  39 &  4.15 $\pm$\  0.04 &  1.23 $\pm$\  0.04 &  0.16 $\pm$\  0.03 & 236, 28 &  1.11 $\pm$\  0.08 &       SARG &  no \\
\object{             HD10697} & 5653 $\pm$\  21 &  4.01 $\pm$\  0.03 &  1.06 $\pm$\  0.03 &  0.13 $\pm$\  0.02 & 234, 33 &  1.17 $\pm$\  0.08 &     SOPHIE & yes \\
\object{            HD108874} & 5590 $\pm$\  60 &  4.34 $\pm$\  0.09 &  0.91 $\pm$\  0.09 &  0.22 $\pm$\  0.04 & 179, 22 &  1.00 $\pm$\  0.08 &        UES & yes \\
\object{            HD109749} & 5885 $\pm$\  34 &  4.30 $\pm$\  0.05 &  1.05 $\pm$\  0.04 &  0.30 $\pm$\  0.03 & 236, 33 &  1.15 $\pm$\  0.08 &    CORALIE &  no \\
\object{            HD109749} & 5881 $\pm$\  22 &  4.28 $\pm$\  0.03 &  1.11 $\pm$\  0.03 &  0.26 $\pm$\  0.02 & 198, 25 &  1.15 $\pm$\  0.08 &       UVES & yes \\
\object{            HD114762} & 5869 $\pm$\  34 &  4.31 $\pm$\  0.03 &  1.03 $\pm$\  0.06 & -0.65 $\pm$\  0.03 & 201, 30 &  0.88 $\pm$\  0.06 &       FIES &  no \\
\object{            HD114762} & 5961 $\pm$\  26 &  4.43 $\pm$\  0.03 &  1.35 $\pm$\  0.06 & -0.65 $\pm$\  0.02 & 158, 21 &  0.87 $\pm$\  0.06 &       UVES & yes \\
\object{             HD11964} & 5360 $\pm$\  26 &  3.86 $\pm$\  0.06 &  1.14 $\pm$\  0.03 &  0.10 $\pm$\  0.02 & 233, 29 &  1.15 $\pm$\  0.09 &      FEROS &  no \\
\object{             HD11964} & 5326 $\pm$\  19 &  3.87 $\pm$\  0.04 &  0.92 $\pm$\  0.03 &  0.10 $\pm$\  0.01 & 239, 32 &  1.12 $\pm$\  0.08 &      HARPS & yes \\
\object{             HD12661} & 5775 $\pm$\  33 &  4.36 $\pm$\  0.06 &  1.09 $\pm$\  0.04 &  0.40 $\pm$\  0.03 & 245, 33 &  1.12 $\pm$\  0.08 &      FEROS & yes \\
\object{             HD12661} & 5749 $\pm$\  37 &  4.30 $\pm$\  0.07 &  1.10 $\pm$\  0.04 &  0.36 $\pm$\  0.03 & 235, 30 &  1.12 $\pm$\  0.08 &        UES &  no \\
\object{            HD142415} & 5977 $\pm$\  27 &  4.52 $\pm$\  0.05 &  1.08 $\pm$\  0.04 &  0.17 $\pm$\  0.02 & 249, 33 &  1.06 $\pm$\  0.07 &      FEROS & yes \\
\object{            HD149143} & 5971 $\pm$\  39 &  4.21 $\pm$\  0.06 &  1.25 $\pm$\  0.04 &  0.35 $\pm$\  0.03 & 222, 28 &  1.26 $\pm$\  0.09 &    CORALIE &  no \\
\object{            HD149143} & 5950 $\pm$\  21 &  4.21 $\pm$\  0.04 &  1.25 $\pm$\  0.03 &  0.32 $\pm$\  0.02 & 246, 32 &  1.24 $\pm$\  0.08 &   ESPADONS & yes \\
\object{            HD150706} & 5928 $\pm$\  22 &  4.47 $\pm$\  0.03 &  0.97 $\pm$\  0.03 & -0.03 $\pm$\  0.02 & 247, 29 &  1.00 $\pm$\  0.07 &     SOPHIE & yes \\
\object{            HD154857} & 5567 $\pm$\  20 &  3.90 $\pm$\  0.03 &  1.15 $\pm$\  0.03 & -0.24 $\pm$\  0.02 & 234, 25 &  1.08 $\pm$\  0.08 &      FEROS &  no \\
\object{            HD154857} & 5560 $\pm$\  12 &  3.90 $\pm$\  0.02 &  1.13 $\pm$\  0.02 & -0.26 $\pm$\  0.01 & 202, 23 &  1.07 $\pm$\  0.07 &       UVES & yes \\
\object{            HD168443} & 5590 $\pm$\  17 &  4.11 $\pm$\  0.03 &  1.05 $\pm$\  0.02 &  0.06 $\pm$\  0.01 & 247, 31 &  1.06 $\pm$\  0.07 &      HARPS & yes \\
\object{           HD178911B} & 5644 $\pm$\  32 &  4.38 $\pm$\  0.04 &  0.80 $\pm$\  0.06 &  0.20 $\pm$\  0.03 & 239, 34 &  1.00 $\pm$\  0.07 &     SOPHIE & yes \\
\object{            HD183263} & 5938 $\pm$\  22 &  4.32 $\pm$\  0.04 &  1.18 $\pm$\  0.03 &  0.30 $\pm$\  0.02 & 229, 25 &  1.17 $\pm$\  0.08 &      FEROS &  no \\
\object{            HD183263} & 5986 $\pm$\  21 &  4.39 $\pm$\  0.04 &  1.11 $\pm$\  0.03 &  0.32 $\pm$\  0.02 & 247, 33 &  1.16 $\pm$\  0.08 &      HARPS & yes \\
\object{            HD185269} & 5998 $\pm$\  43 &  3.98 $\pm$\  0.06 &  1.42 $\pm$\  0.05 &  0.14 $\pm$\  0.03 & 228, 31 &  1.35 $\pm$\  0.10 &     ELODIE & yes \\
\object{            HD187123} & 5835 $\pm$\  18 &  4.38 $\pm$\  0.03 &  1.04 $\pm$\  0.02 &  0.13 $\pm$\  0.01 & 249, 33 &  1.04 $\pm$\  0.07 &   ESPADONS & yes \\
\object{            HD187123} & 5877 $\pm$\  27 &  4.50 $\pm$\  0.04 &  1.05 $\pm$\  0.04 &  0.16 $\pm$\  0.02 & 231, 27 &  1.03 $\pm$\  0.07 &       SARG &  no \\
\object{            HD188015} & 5726 $\pm$\  28 &  4.35 $\pm$\  0.06 &  1.05 $\pm$\  0.04 &  0.27 $\pm$\  0.02 & 224, 23 &  1.06 $\pm$\  0.08 &      FEROS & yes \\
\object{            HD190228} & 5301 $\pm$\  16 &  3.79 $\pm$\  0.02 &  0.96 $\pm$\  0.02 & -0.24 $\pm$\  0.01 & 202, 23 &  1.05 $\pm$\  0.07 &       UVES & yes \\
\object{            HD190360} & 5604 $\pm$\  28 &  4.31 $\pm$\  0.05 &  0.96 $\pm$\  0.04 &  0.23 $\pm$\  0.02 & 241, 32 &  1.02 $\pm$\  0.07 &   ESPADONS & yes \\
\object{            HD190360} & 5620 $\pm$\  39 &  4.28 $\pm$\  0.07 &  1.02 $\pm$\  0.05 &  0.26 $\pm$\  0.03 & 183, 21 &  1.05 $\pm$\  0.08 &        UES &  no \\
\object{            HD195019} & 5790 $\pm$\  20 &  4.19 $\pm$\  0.04 &  1.06 $\pm$\  0.03 &  0.09 $\pm$\  0.02 & 213, 28 &  1.10 $\pm$\  0.08 &    CORALIE &  no \\
\object{            HD195019} & 5800 $\pm$\  14 &  4.21 $\pm$\  0.03 &  1.08 $\pm$\  0.02 &  0.07 $\pm$\  0.01 & 201, 24 &  1.08 $\pm$\  0.07 &       UVES & yes \\
\object{           HD196885A} & 6308 $\pm$\  55 &  4.42 $\pm$\  0.05 &  1.59 $\pm$\  0.07 &  0.27 $\pm$\  0.04 & 198, 22 &  1.26 $\pm$\  0.09 &       SARG & yes \\
\object{              HD2039} & 5948 $\pm$\  39 &  4.42 $\pm$\  0.08 &  1.01 $\pm$\  0.05 &  0.37 $\pm$\  0.03 & 234, 33 &  1.15 $\pm$\  0.08 &    CORALIE &  no \\
\object{              HD2039} & 5968 $\pm$\  22 &  4.42 $\pm$\  0.03 &  1.10 $\pm$\  0.03 &  0.33 $\pm$\  0.02 & 196, 24 &  1.15 $\pm$\  0.08 &       UVES & yes \\
\object{            HD216437} & 5860 $\pm$\  19 &  4.21 $\pm$\  0.03 &  1.24 $\pm$\  0.02 &  0.24 $\pm$\  0.02 & 199, 25 &  1.17 $\pm$\  0.08 &       UVES & yes \\
\object{            HD217107} & 5657 $\pm$\  34 &  4.29 $\pm$\  0.07 &  1.17 $\pm$\  0.04 &  0.34 $\pm$\  0.03 & 232, 26 &  1.08 $\pm$\  0.08 &      FEROS &  no \\
\object{            HD217107} & 5647 $\pm$\  30 &  4.28 $\pm$\  0.05 &  0.96 $\pm$\  0.04 &  0.34 $\pm$\  0.02 & 188, 24 &  1.09 $\pm$\  0.08 &       UVES & yes \\
\object{            HD219828} & 5888 $\pm$\  14 &  4.20 $\pm$\  0.02 &  1.17 $\pm$\  0.01 &  0.18 $\pm$\  0.01 & 242, 32 &  1.16 $\pm$\  0.08 &      HARPS & yes \\
\object{             HD23596} & 6099 $\pm$\  33 &  4.30 $\pm$\  0.06 &  1.29 $\pm$\  0.04 &  0.31 $\pm$\  0.03 & 230, 28 &  1.25 $\pm$\  0.09 &        UES & yes \\
\object{             HD30562} & 5957 $\pm$\  31 &  4.22 $\pm$\  0.04 &  1.23 $\pm$\  0.03 &  0.26 $\pm$\  0.03 & 226, 32 &  1.21 $\pm$\  0.08 &    CORALIE &  no \\
\object{             HD30562} & 5945 $\pm$\  41 &  4.10 $\pm$\  0.07 &  1.25 $\pm$\  0.05 &  0.27 $\pm$\  0.03 & 209, 25 &  1.30 $\pm$\  0.09 &     ELODIE &  no \\
\object{             HD30562} & 5963 $\pm$\  24 &  4.24 $\pm$\  0.05 &  1.28 $\pm$\  0.03 &  0.28 $\pm$\  0.02 & 228, 27 &  1.21 $\pm$\  0.08 &      FEROS & yes \\
\object{             HD34445} & 5840 $\pm$\  23 &  4.19 $\pm$\  0.04 &  1.21 $\pm$\  0.03 &  0.13 $\pm$\  0.02 & 224, 27 &  1.13 $\pm$\  0.08 &    CORALIE & yes \\
\object{             HD37124} & 5503 $\pm$\  32 &  4.42 $\pm$\  0.05 &  0.62 $\pm$\  0.07 & -0.41 $\pm$\  0.03 & 238, 30 &  0.82 $\pm$\  0.06 &       FIES &  no \\
\object{             HD37124} & 5510 $\pm$\  16 &  4.39 $\pm$\  0.03 &  0.84 $\pm$\  0.03 & -0.43 $\pm$\  0.01 & 190, 21 &  0.82 $\pm$\  0.06 &       UVES & yes \\
\object{             HD37605} & 5450 $\pm$\  46 &  4.31 $\pm$\  0.10 &  1.09 $\pm$\  0.07 &  0.28 $\pm$\  0.03 & 232, 31 &  0.98 $\pm$\  0.08 &      FEROS & yes \\
\object{             HD38529} & 5628 $\pm$\  34 &  3.81 $\pm$\  0.07 &  1.36 $\pm$\  0.04 &  0.36 $\pm$\  0.03 & 222, 26 &  1.40 $\pm$\  0.11 &      FEROS &  no \\
\object{             HD38529} & 5653 $\pm$\  29 &  3.83 $\pm$\  0.05 &  1.29 $\pm$\  0.03 &  0.37 $\pm$\  0.02 & 194, 25 &  1.40 $\pm$\  0.10 &       UVES & yes \\
\object{            HD41004A} & 5255 $\pm$\  52 &  4.34 $\pm$\  0.11 &  0.97 $\pm$\  0.08 &  0.15 $\pm$\  0.03 & 222, 30 &  0.89 $\pm$\  0.07 &      FEROS & yes \\
\object{              HD4203} & 5652 $\pm$\  33 &  4.14 $\pm$\  0.08 &  1.19 $\pm$\  0.04 &  0.39 $\pm$\  0.03 & 219, 25 &  1.18 $\pm$\  0.09 &      FEROS &  no \\
\object{              HD4203} & 5656 $\pm$\  38 &  4.10 $\pm$\  0.06 &  1.13 $\pm$\  0.04 &  0.38 $\pm$\  0.03 & 189, 24 &  1.21 $\pm$\  0.09 &       UVES & yes \\
\object{             HD43691} & 6258 $\pm$\  41 &  4.31 $\pm$\  0.07 &  1.45 $\pm$\  0.04 &  0.32 $\pm$\  0.03 & 221, 29 &  1.32 $\pm$\  0.09 &     SOPHIE & yes \\
\object{             HD46375} & 5305 $\pm$\  50 &  4.39 $\pm$\  0.09 &  0.87 $\pm$\  0.08 &  0.23 $\pm$\  0.03 & 232, 34 &  0.90 $\pm$\  0.07 &   ESPADONS & yes \\
\object{             HD46375} & 5290 $\pm$\  65 &  4.26 $\pm$\  0.12 &  1.09 $\pm$\  0.08 &  0.17 $\pm$\  0.04 & 229, 27 &  0.93 $\pm$\  0.08 &        UES &  no \\
\object{             HD49674} & 5662 $\pm$\  28 &  4.42 $\pm$\  0.05 &  0.93 $\pm$\  0.04 &  0.30 $\pm$\  0.02 & 242, 30 &  1.03 $\pm$\  0.07 &   ESPADONS & yes \\
\object{             HD49674} & 5674 $\pm$\  53 &  4.49 $\pm$\  0.07 &  0.86 $\pm$\  0.09 &  0.34 $\pm$\  0.04 & 216, 27 &  1.02 $\pm$\  0.07 &       SARG &  no \\
\object{             HD49674} & 5628 $\pm$\  33 &  4.39 $\pm$\  0.05 &  0.82 $\pm$\  0.05 &  0.27 $\pm$\  0.02 & 242, 33 &  1.01 $\pm$\  0.07 &     SOPHIE &  no \\
\object{             HD50499} & 6043 $\pm$\  34 &  4.23 $\pm$\  0.04 &  1.31 $\pm$\  0.04 &  0.33 $\pm$\  0.03 & 233, 30 &  1.27 $\pm$\  0.09 &    CORALIE &  no \\
\object{             HD50499} & 6077 $\pm$\  19 &  4.36 $\pm$\  0.04 &  1.27 $\pm$\  0.02 &  0.34 $\pm$\  0.01 & 248, 33 &  1.22 $\pm$\  0.08 &      HARPS & yes \\
\object{             HD50554} & 6061 $\pm$\  23 &  4.47 $\pm$\  0.03 &  1.18 $\pm$\  0.03 &  0.03 $\pm$\  0.02 & 204, 24 &  1.06 $\pm$\  0.07 &       UVES &  no \\
\object{              HD6434} & 5721 $\pm$\  45 &  4.29 $\pm$\  0.04 &  0.82 $\pm$\  0.08 & -0.53 $\pm$\  0.04 & 154, 18 &  0.87 $\pm$\  0.06 &      FEROS & yes \\
\object{             HD68988} & 5946 $\pm$\  64 &  4.39 $\pm$\  0.12 &  1.35 $\pm$\  0.08 &  0.34 $\pm$\  0.05 & 191, 23 &  1.16 $\pm$\  0.09 &       SARG & yes \\
\object{             HD73526} & 5629 $\pm$\  32 &  4.08 $\pm$\  0.06 &  1.12 $\pm$\  0.04 &  0.26 $\pm$\  0.03 & 242, 31 &  1.16 $\pm$\  0.08 &      FEROS &  no \\
\object{             HD73526} & 5661 $\pm$\  25 &  4.15 $\pm$\  0.04 &  1.10 $\pm$\  0.03 &  0.26 $\pm$\  0.02 & 203, 24 &  1.13 $\pm$\  0.08 &       UVES & yes \\
\object{             HD74156} & 6108 $\pm$\  30 &  4.37 $\pm$\  0.04 &  1.29 $\pm$\  0.04 &  0.17 $\pm$\  0.02 & 218, 26 &  1.16 $\pm$\  0.08 &      FEROS &  no \\
\object{             HD74156} & 6065 $\pm$\  21 &  4.25 $\pm$\  0.03 &  1.28 $\pm$\  0.03 &  0.13 $\pm$\  0.02 & 206, 24 &  1.18 $\pm$\  0.08 &       UVES & yes \\
\object{             HD76700} & 5645 $\pm$\  29 &  4.14 $\pm$\  0.05 &  1.12 $\pm$\  0.03 &  0.35 $\pm$\  0.02 & 219, 22 &  1.16 $\pm$\  0.08 &      FEROS &  no \\
\object{             HD76700} & 5689 $\pm$\  29 &  4.18 $\pm$\  0.05 &  1.06 $\pm$\  0.04 &  0.37 $\pm$\  0.02 & 199, 25 &  1.16 $\pm$\  0.08 &       UVES & yes \\
\object{             HD80606} & 5542 $\pm$\  36 &  4.28 $\pm$\  0.06 &  0.79 $\pm$\  0.05 &  0.27 $\pm$\  0.03 & 237, 34 &  1.02 $\pm$\  0.07 &     SOPHIE & yes \\
\object{             HD80606} & 5582 $\pm$\  81 &  4.60 $\pm$\  0.15 &  1.03 $\pm$\  0.12 &  0.25 $\pm$\  0.06 & 204, 28 &  0.95 $\pm$\  0.08 &        UES &  no \\
\object{             HD81040} & 5684 $\pm$\  30 &  4.38 $\pm$\  0.04 &  0.91 $\pm$\  0.05 & -0.08 $\pm$\  0.02 & 232, 33 &  0.94 $\pm$\  0.07 &    CORALIE &  no \\
\object{             HD81040} & 5778 $\pm$\  23 &  4.54 $\pm$\  0.03 &  0.88 $\pm$\  0.04 & -0.03 $\pm$\  0.02 & 246, 31 &  0.94 $\pm$\  0.07 &       FIES & yes \\
\object{              HD8574} & 6070 $\pm$\  41 &  4.26 $\pm$\  0.05 &  1.15 $\pm$\  0.05 &  0.06 $\pm$\  0.03 & 219, 26 &  1.16 $\pm$\  0.08 &       SARG & yes \\
\object{             HD88133} & 5516 $\pm$\  46 &  4.06 $\pm$\  0.07 &  1.05 $\pm$\  0.06 &  0.40 $\pm$\  0.04 & 139, 17 &  1.18 $\pm$\  0.09 &       UVES & yes \\
\object{             HD89307} & 5992 $\pm$\  17 &  4.49 $\pm$\  0.03 &  1.06 $\pm$\  0.03 & -0.09 $\pm$\  0.01 & 240, 30 &  1.00 $\pm$\  0.07 &   ESPADONS & yes \\
\object{             HD89307} & 5961 $\pm$\  18 &  4.46 $\pm$\  0.03 &  1.00 $\pm$\  0.03 & -0.12 $\pm$\  0.01 & 243, 31 &  0.98 $\pm$\  0.07 &     SOPHIE &  no \\
\object{             HD89744} & 6381 $\pm$\  43 &  4.27 $\pm$\  0.05 &  1.70 $\pm$\  0.05 &  0.30 $\pm$\  0.03 & 145, 21 &  1.37 $\pm$\  0.09 &       SARG & yes \\
\object{               muAra} & 5784 $\pm$\  21 &  4.22 $\pm$\  0.04 &  1.07 $\pm$\  0.03 &  0.30 $\pm$\  0.02 & 242, 32 &  1.15 $\pm$\  0.08 &      HARPS & yes \\
\object{               muAra} & 5799 $\pm$\  24 &  4.29 $\pm$\  0.04 &  1.06 $\pm$\  0.03 &  0.31 $\pm$\  0.02 & 200, 24 &  1.13 $\pm$\  0.08 &       UVES &  no \\
\object{          OGLE-TR-10} & 6172 $\pm$\  73 &  4.58 $\pm$\  0.07 &  1.53 $\pm$\  0.10 &  0.37 $\pm$\  0.06 & 183, 23 &  1.19 $\pm$\  0.08 &       UVES & yes \\
\object{         OGLE-TR-132} & 6180 $\pm$\  43 &  4.35 $\pm$\  0.05 &  1.41 $\pm$\  0.05 &  0.30 $\pm$\  0.03 & 177, 18 &  1.25 $\pm$\  0.09 &       UVES & yes \\
\object{          OGLE-TR-56} & 6154 $\pm$\  62 &  4.25 $\pm$\  0.10 &  1.47 $\pm$\  0.07 &  0.28 $\pm$\  0.05 & 196, 25 &  1.30 $\pm$\  0.10 &       UVES & yes \\
\object{              tauBoo} & 6617 $\pm$\  80 &  4.54 $\pm$\  0.10 &  1.82 $\pm$\  0.10 &  0.40 $\pm$\  0.06 & 141, 19 &  1.38 $\pm$\  0.10 &   ESPADONS &  no \\
\object{              tauBoo} & 6659 $\pm$\  84 &  4.57 $\pm$\  0.10 &  1.94 $\pm$\  0.10 &  0.40 $\pm$\  0.06 & 140, 21 &  1.38 $\pm$\  0.10 &       UVES & yes \\
\object{              TrES-1} & 5285 $\pm$\  45 &  4.37 $\pm$\  0.09 &  0.96 $\pm$\  0.08 &  0.08 $\pm$\  0.03 & 188, 24 &  0.87 $\pm$\  0.07 &       UVES & yes \\

\end{longtable}
}

\begin{acknowledgements}

This work was financed by FEDER - Fundo Europeu de Desenvolvimento Regional funds through the COMPETE 2020 - Operacional Programme for Competitiveness and Internationalisation (POCI), and by Portuguese funds through FCT - Funda\c{c}\~ao para a Ci\^encia e a Tecnologia in the framework of 
the projects POCI-01-0145-FEDER-028953, PTDC/FIS-AST/7073/2014 \& POCI-01-0145-FEDER-016880, and  PTDC/FIS-AST/1526/2014 \& POCI-01-0145-FEDER-016886. P.F., S.C.C.B., N.C.S., S.G.S, V.A. and E.D.M. acknowledge support from FCT through Investigador FCT contracts nr. IF/01037/2013CP1191/CT0001, IF/01312/2014/CP1215/CT0004, IF/00169/2012/CP0150/CT0002, IF/00028/2014/CP1215/CT0002, IF/00650/2015/CP1273/CT0001, and IF/00849/2015/CP1273/CT0003. S.G.S. further acknowledges support from FCT in the form of an exploratory project of reference IF/00028/2014/CP1215/CT0002. ACSF is supported by grant 234989/2014-9 from CNPq (Brazil). J.P.F. and acknowledges support from FCT through the grants SFRH/BD/93848/2013 and SFRH/BPD/87857/2012. B.R.-A. acknowledges the support from CONICYT PAI/Concurso Nacional Inserción en la Academia, Convocatoria 2015 79150050.

\end{acknowledgements}

\bibliographystyle{aa}
\bibliography{sousa_bibliography}

\end{document}